\documentclass[twoside]{IEEEtran}

\normalsize
\usepackage[utf8]{inputenc}
\usepackage[cmex10]{amsmath}
\usepackage{algorithmic}
\usepackage{array}
\usepackage{mdwmath}
\usepackage{mdwtab}

\usepackage{xcolor}
\usepackage{graphicx}
\usepackage{amsmath}
\usepackage{booktabs}

\usepackage{color}
\usepackage{eqparbox}

\usepackage[utf8]{inputenc} 
\usepackage[T1]{fontenc}    
\usepackage{booktabs}       
\usepackage{amsfonts}       
\usepackage{nicefrac}       
\usepackage{microtype}      
\usepackage{comment}

\usepackage[utf8]{inputenc} 
\usepackage[T1]{fontenc}    
\usepackage{hyperref}       
\usepackage{url}                
\usepackage{booktabs}       
\usepackage{amsfonts}       
\usepackage{nicefrac}       
\usepackage{microtype}      

\usepackage{comment}
\usepackage{amsmath}
\usepackage{color}
\usepackage{xcolor}

\usepackage{graphicx}

\def\ol{\color{black}}

\usepackage[letterpaper,
            left=0.6in,right=0.6in,
            top=0.7in,bottom=0.9in,%
            footskip=.25in]{geometry}

\begin{document}
%
\title{LEARN Codes: Inventing Low-Latency Codes via Recurrent Neural Networks}
\author{Yihan Jiang,
\thanks{Y. Jiang and S. Kannan are with Department of Electrical and Computer Engineering (ECE) at University of Washington (UW), Seattle, USA. Email: \texttt{yihanrogerjiang@gmail.com} (Y. Jiang), \texttt{ksreeram@uw.edu} (S. Kannan). H. Kim is with Samsung AI Center Cambridge, Cambridge, United Kingdom, Email: \texttt{hkim1505@gmail.com}. H. Asnani is with the School of Technology and Computer Science (STCS) at Tata Institute of Fundamental Research (TIFR), Mumbai, India, Email: \texttt{himanshu.asnani@tifr.res.in}. 
S. Oh is with the Allen School of Computer Science $\&$ Engineering at University of Washington (UW), Seattle, USA, Email: \texttt{sewoong@cs.washington.edu}. P. Viswanath is with the Coordinated Science Lab (CSL) and the Department of Electrical Engineering at University of Illinois at Urbana Champaign (UIUC), Email: \texttt{pramodv@illinois.edu}.\newline
This paper is an extended version of work appeared in the 53rd IEEE International Conference on Communications (ICC 2019).\newline
} 
Hyeji Kim, Himanshu Asnani, Sreeram Kannan, Sewoong Oh, Pramod Viswanath}

%

\maketitle

\begin{abstract}
Designing channel codes under low-latency constraints is one of the most demanding requirements in 5G standards. However, a sharp characterization of the performance of traditional codes is available only in the large block-length limit. Guided by such asymptotic analysis, code designs require large block lengths as well as latency to achieve the desired error rate. Tail-biting convolutional codes and other recent state-of-the-art short block codes, while promising reduced latency, are neither robust to channel-mismatch nor adaptive to varying channel conditions.  When the codes designed for one channel (e.g.,~Additive White Gaussian Noise (AWGN) channel) are used for another (e.g.,~non-AWGN channels), heuristics are necessary to achieve non-trivial performance. 


In this paper, we first propose an end-to-end learned neural code, obtained by jointly designing a Recurrent Neural Network (RNN) based encoder and decoder. This code outperforms canonical convolutional code under block settings. 
%
We then leverage this experience to propose a new class of codes under low-latency constraints, which we call \textit{Low-latency Efficient Adaptive Robust Neural (LEARN) codes}. These codes outperform state-of-the-art low-latency codes and exhibit robustness and adaptivity properties. LEARN codes show the potential to design new versatile and universal codes for future communications via tools of modern deep learning coupled with communication engineering insights.


\end{abstract}


%
\IEEEpeerreviewmaketitle

\section{Introduction}\label{intro}
Reliable channel codes have had a huge impact for communications in the modern information age.
Since its inception in \cite{shannon2001mathematical} that was powered by the mathematical insights of information theory and principles of modern engineering, several capacity-achieving codes such as Polar, Turbo and LDPC codes \cite{arikan2008performance}\cite{berrou1993near}\cite{mackay1996near} have come close to the Shannon limit when operated at large block lengths under Additive White Gaussian Noise (AWGN) channels. These codes were successfully adopted and applied in LTE and 5G data planes~\cite{richardson2008modern}. Since 5G is under intensive development, designing codes that have features such as \emph{low latency}, \emph{robustness}, and \emph{adaptivity} has become increasingly important. 

\subsection{Motivation}
An Ultra-Reliable Low Latency Communication (URLLC) code \cite{sybis2016channel} requires minimal delay constraints, thereby enabling scenarios such as vehicular communication, virtual reality, and remote surgery. When considering low latency requirements, it is instructive to observe the interplay of three different types of delays: \emph{processing delays}, \emph{propagation delays}, and \emph{structural delays}. Processing and propagation delays are affected mostly by computing resources and varying environments \cite{rachinger2015comparison}. Low-latency channel coding, the focus of this paper, aims to improve the structural delay caused by the encoder and/or decoder. Encoder structural delay refers to the delay between the encoder's receiving the information bit and sending it out. Decoder structural delay refers to the delay between the decoder's receiving and decoding bits from the channel. Traditional AWGN capacity-achieving codes, such as LDPC and Turbo codes with small block lengths, show poor performance for URLLC requirements \cite{rachinger2015comparison}, \cite{maiya2012low}. There has also been recent interest in establishing theoretical limits and bounds on the reliability of codes at small to medium block lengths \cite{polyanskiy2010channel}.\\

We note that latency is proportional to block length when using a block code;
the decoder waits until it receives the entire (noisy) codeword to start the decoding.
However, when using a convolutional code, latency is given by the decoding window length. Thus, there is an inherent difference between block codes and convolutional codes when considering latency. Since the latter incorporates locality in encoding, they can also be locally decoded. While convolutional codes with small constraint lengths are not capacity achieving, they can possibly be optimal under the low latency constraint. Indeed, this possibility was raised by \cite{rachinger2015comparison} and observed in \cite{liva1}, \cite{liva2}, \cite{liva3}. 
In this work, we advance this hypothesis by showing that we can invent codes similar to convolutional codes that outperform handcrafted codes in the low latency regime. While convolutional codes are the state-of-the-art in this regime, in the moderate latency regime, Extended Bose-Chaudhuri-Hocquenghem codes (eBCH) also perform well \cite{shirvanimoghaddam2018short}. \\

In addition, low latency constraint channel coding must take channel effects into account in fading channels because pilot bits used for accurate channel estimation increase latency \cite{rachinger2015comparison}. 
This calls for incorporating both robustness and adaptivity as desired features for URLLC codes. 
\textit{Robustness} refers to the ability to perform with acceptable degradation without retraining when model mismatches occur; \textit{adaptivity} refers to the ability to learn to adapt to different channel models with retraining.  Current and traditional channel coding systems require heuristics to compensate for channel effects, which leads to sub-optimality for model mismatches~\cite{li2013ofdma}. In general, channels without clean mathematical analysis lack the theory of an optimal communication algorithm and thus rely on sub-optimal heuristics \cite{safavi2015impact}. 
In short, current channel coding schemes fail to deliver under the challenges of low latency, robustness, and adaptivity. 

\subsection{Channel Coding Inspired by Deep Learning: Prior Art}
On the past decade, advances in \emph{deep learning} (DL) have greatly benefited engineering fields such as computer vision, natural language processing, and gaming technology \cite{goodfellow2016deep}. 
This has generated recent excitement about applying DL methods to communication system design  \cite{o2016learning}\cite{o2017introduction}. These methods have typically been successful in settings where there is a significant \emph{model-deficiency}, i.e., the observed data cannot be well described by a clean mathematical model. Thus, many initial proposals applying DL to communication systems have also focused on problems where there is model uncertainty due to the lack of, say, channel knowledge \cite{more_Nariman, dorner2017deep}. In developing codes for the AWGN channel under low latency constraints, there is no model deficiency since the channel is well-defined mathematically and simple to describe. However, the main challenge is that optimal codes and decoding algorithms are not known; we refer to this as \emph{algorithm deficiency}. 

\begin{figure}[!h] 
\centering
\includegraphics[width=0.47\textwidth]{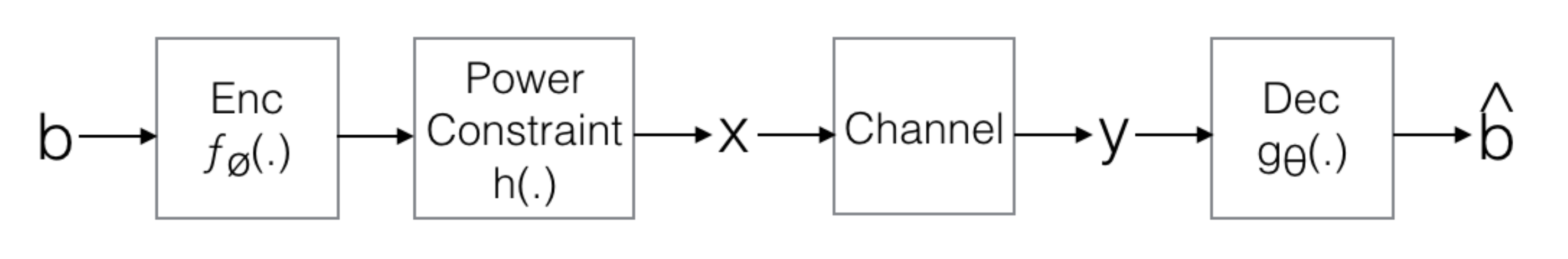}\ \ \ 
\caption{Channel AE block diagram.}\label{ChannelAE}
\centering
\end{figure}

For \emph{algorithm-deficit} problems, the following two categories of work apply deep learning to communication systems: (1) designing a neural network decoder (also known as neural decoder) for a given canonical encoder such as LDPC or Turbo codes, and (2) jointly designing both the neural network encoder and decoder, referred to as a Channel Autoencoder (also known as Channel AE) \cite{o2016learning} (as illustrated in Figure \ref{ChannelAE}). 

Neural decoders show promising performance by mimicking and modifying existing optimal decoding algorithms. Learnable Belief Propagation (BP) decoders for BCH and High-Density Parity-Check (HDPC) codes have been proposed in \cite{nachmani2016learning} and \cite{nachmani2018deep}. Polar decoding via neural BP is proposed in \cite{gruber2017deep} and \cite{cammerer2017scaling}. Since mimicking learnable Tanner graphs requires a fully connected neural network, generalizing to longer block lengths is prohibitive. Capacity-achieving performance for Turbo code under an AWGN channel is achieved via Recurrent Neural Networks (RNNs) for arbitrary block lengths \cite{kim2018communication}. 
The joint design of neural code (encoders) and decoders via a Channel AE, relevant to the problem under consideration in this paper, has witnessed scant progress. Deep autoencoders 
have been successfully applied for various problems, such as dimensionality reduction, representation learning, and graph generation \cite{hinton2006reducing}. However, Channel AE significantly differs from typical deep autoencoder models in the following two aspects, making its design and training highly challenging:

\begin{enumerate}
\item The number of possible message bits $b$ grows exponentially with respect to the block length. Hence, Channel AE must generalize to unseen messages with capacity-restricted encoders and decoders \cite{gruber2017deep}. 
\item Channel models add noise between the encoder and decoder, and the encoder needs to satisfy power constraints, thus requiring high robustness in the code. 
\end{enumerate}


	For Channel AE training, \cite{o2016learning} and \cite{o2017introduction} introduce learning tricks emphasizing both channel coding and modulations. Learning Channel AE without a channel gradient is shown in \cite{aoudia2018end}. Modulation gain is reported in \cite{felix2018ofdm}. Beyond AWGN and fading channels, \cite{kim2018deepcode} extended RNN to design a code for the feedback channel, which outperforms existing state-of-the-art codes \cite{SK, CL}. Extending Channel AE to MIMO settings is reported in \cite{o2017deep}. 
	Despite the successes, existing research on Channel AE under canonical channels is currently restricted to very short block lengths (for example, achieving the same performance as a rate 4/7 Hamming code with 4 information bits). 
Furthermore, existing works do not focus on the low latency, robustness, and adaptivity requirements. 

With this backdrop, this paper poses the following fundamental question:  \textit{Can we improve the Channel AE design to construct new codes that comply with low latency requirements?}

We answer this in affirmative, as described next. 

\subsection{Our Contribution}
Our primary goal is to design a low latency code under extremely low structural-latency requirements. 
As noted, convolutional codes outperform block codes given  low structural latency \cite{rachinger2015comparison} \cite{liva1} \cite{liva2} \cite{liva3}. An RNN is a constrained neural structure with a natural connection to convolutional codes since the encoded symbol has a locality of memory and is most strongly influenced by the recent past of the input bits.  Furthermore, RNN-based codes have shown a natural generalization across different block lengths in prior work \cite{kim2018communication}\cite{gruber2017deep}. With a carefully designed learnable structure that uses Bidirectional RNN (Bi-RNN) for both encoder and decoder, as well as a novel training methodology developed specifically for the Channel AE model, we demonstrate that our Bi-RNN-based neural code outperforms convolutional code.

We then propose \textit{Low-Latency Efficient Adaptive Robust Neural (LEARN) code}, which applies learnable RNN structures to both the encoder and decoder with an additional low latency constraint. LEARN improves performance under extremely low latency constraints. Ours is the first work we know of that creates an end-to-end design for a neural code that improves performance on the AWGN channel (in any regime). In summary, the contributions of this paper include:

\begin{enumerate}

\item[1.] {\em Outperforming convolutional codes.} We propose a {bi-directional} RNN network structure and a tailored learning methodology for Channel AE that outperform canonical convolutional codes. The proposed training methodology results in smoother training dynamics and better generalization. (Section II)
\item[2.] {\em Improving performance in low latency settings}. We design \textbf{LEARN code} for low latency requirements with specific network designs. LEARN code {\color{black} improves} performance in extremely low latency settings. (Section III). 

\item[3.] {\em Showing robustness and adaptivity}. When channel conditions are varying, LEARN codes demonstrate robustness (ability to work well under unseen channels) and adaptivity (ability to adapt to a new channel with few training symbols), showing an order of magnitude improvement in reliability compared to existing state-of-the-art codes. (Section III) 

\item[4.] {\em Interpreting the neural code}. We provide interpretations to aid in the fundamental understanding of why the jointly trained code works better than canonical codes. (Section IV)

\end{enumerate}


\section{Designing Neural Codes that Outperform Convolutional Codes}\label{earn}
The reliability of neural codes relies heavily on two factors: (1) a network architecture, and (2) a training methodology. 
In this section, we provide guidelines for these two factors and demonstrate that neural codes designed and trained using our guidelines have superior reliability compared to convolutional codes in a block coding setting. 

\subsection{Network Architecture}
The performance (coding gain) of traditional channel codes improves with block length. Recent research on the Channel AE model does not show coding gain for even moderate block lengths \cite{o2016learning}\cite{gruber2017deep} with fully connected neural networks, even with nearly unlimited training examples. 
We argue that a Recurrent Neural Network (RNN) architecture is a more suitable DL structure for Channel AE. For a brief introduction to RNN and its variants -- such as Bidirectional RNN (Bi-RNN), Gated Recurrent Unit (GRU), or Long Short Term Memory (LSTM) -- please refer to Appendix \ref{appendix:rnn_intro}. In this paper, we use the terms GRU and RNN interchangeably.  

\begin{figure*}[!ht] 
\centering
\includegraphics[width=0.8\textwidth]{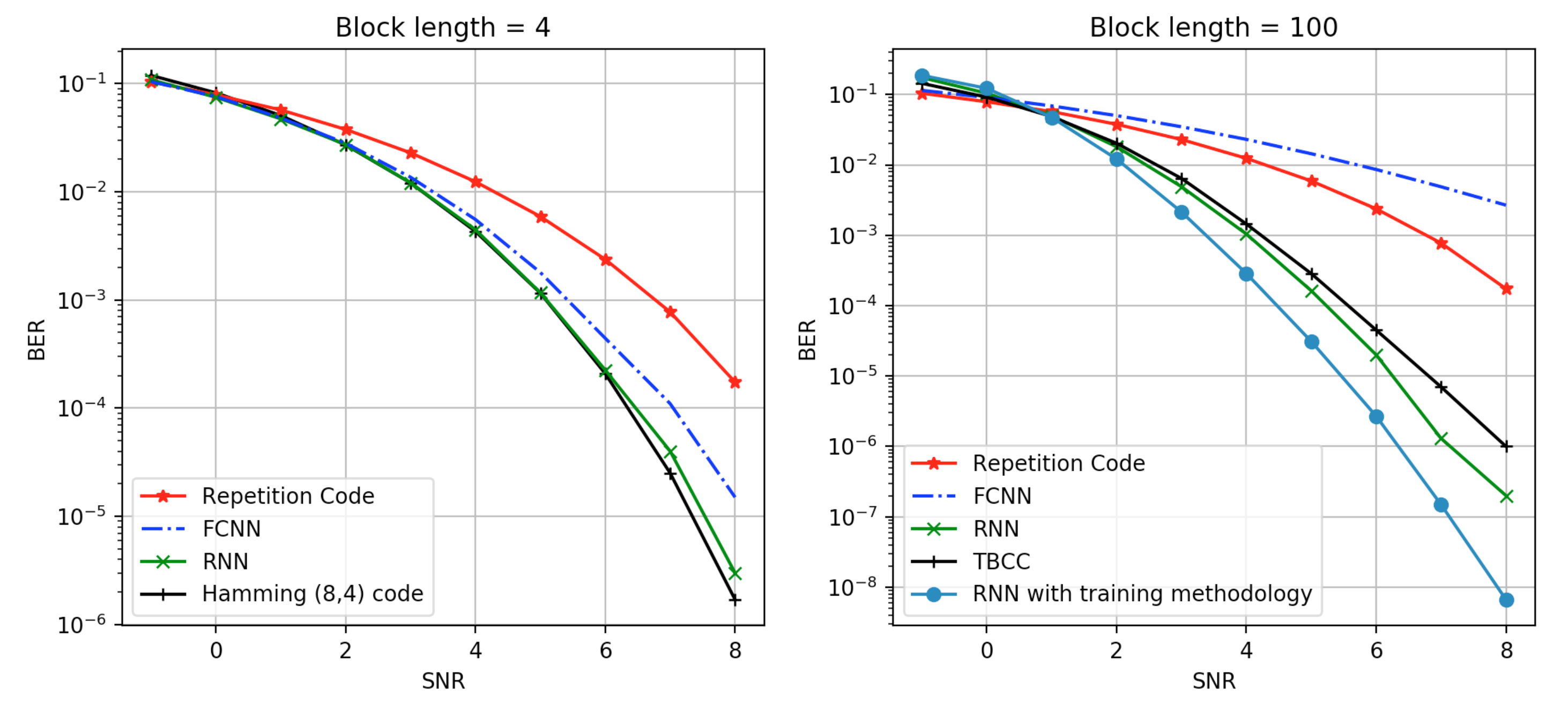}\ \ \ 
\caption{Channel AE performance on code rate 1/2, where block length (number of information bits) is 4 (left) and 100 (right). TBCC is m=2, with $g_{11}=5$, $g_{12}=7$ with feedback $7$.} \label{nn_select}
\end{figure*}

\textbf{RNN-based Encoder and Decoder Design}\\
Our empirical results comparing different Channel AE structures (Figure \ref{nn_select}) show that for long enough block lengths, RNN outperforms 
a Fully Connected Neural Network (FCNN) for Channel AE. The FCNN curve in Figure \ref{nn_select} refers to using FCNN for both encoder and decoder. RNN in Figure \ref{nn_select} refers to using Bi-RNN for both encoder and decoder. The training steps are kept the same for a fair comparison. The repetition code and extended Hamming code performances are shown as a reference for both short and long block lengths. 

Figure \ref{nn_select} (left) shows that for short block lengths (4), the performance of FCNN and RNN are close to each other since enumerating all possible code is not prohibitive. On the other hand, for a longer block length (100), Figure \ref{nn_select} (right) shows that in using FCNN, the Bit Error Rate (BER) is even worse than repetition codes, which shows a failure in generalization; 
RNN outperforms FCNN due to its generalization via parameter sharing and adaptive learnable dependency length. Hence, in this paper, we model the encoder and decoder as RNNs
in order to gain generalization across block lengths. Figure \ref{nn_select} also shows that RNNs with tailored training methodologies outperform simply applying RNN or FCNN for Channel AE; we illustrate this training methodology in Section II.B. 

\textbf{Power Constraint Module}\\
The output of the RNN encoder can take arbitrary values and does not necessarily satisfy the power constraint. To impose the power constraint, we use a power constraint layer after the RNN encoding layer, as shown in Figure \ref{ChannelAE}. 
Power normalization enforces that the output code has unit power by normalizing the output as $E[x^2]=1$. 
More detail is in the Appendix.

\subsection{Training Methodology}
We find in this paper that the following training methods result in a faster learning trajectory and better generalization with the learnable structure discussed above.

  \begin{itemize}
  \item Training with a large batch size
  \item Using Binary Crossentropy (BCE) loss 
  \item Training encoder and decoder separately  
  \item Adding a minimum distance regularizer on encoder. 
  \item Using the Adam optimizer
  \item Adding more capacity (parameters) to the decoder than the encoder
  \end{itemize}

Some of the training methods are not common in deep learning due to the unique structure of Channel AE. Appendix \ref{appendix:training} shows the empirical evidence, where we reason about the better performance of the above training methods.



\subsection{Performance of RNN-Based Channel AE: AWGN Setting}

\begin{figure}[!ht] 
\centering
\includegraphics[width=0.35\textwidth]{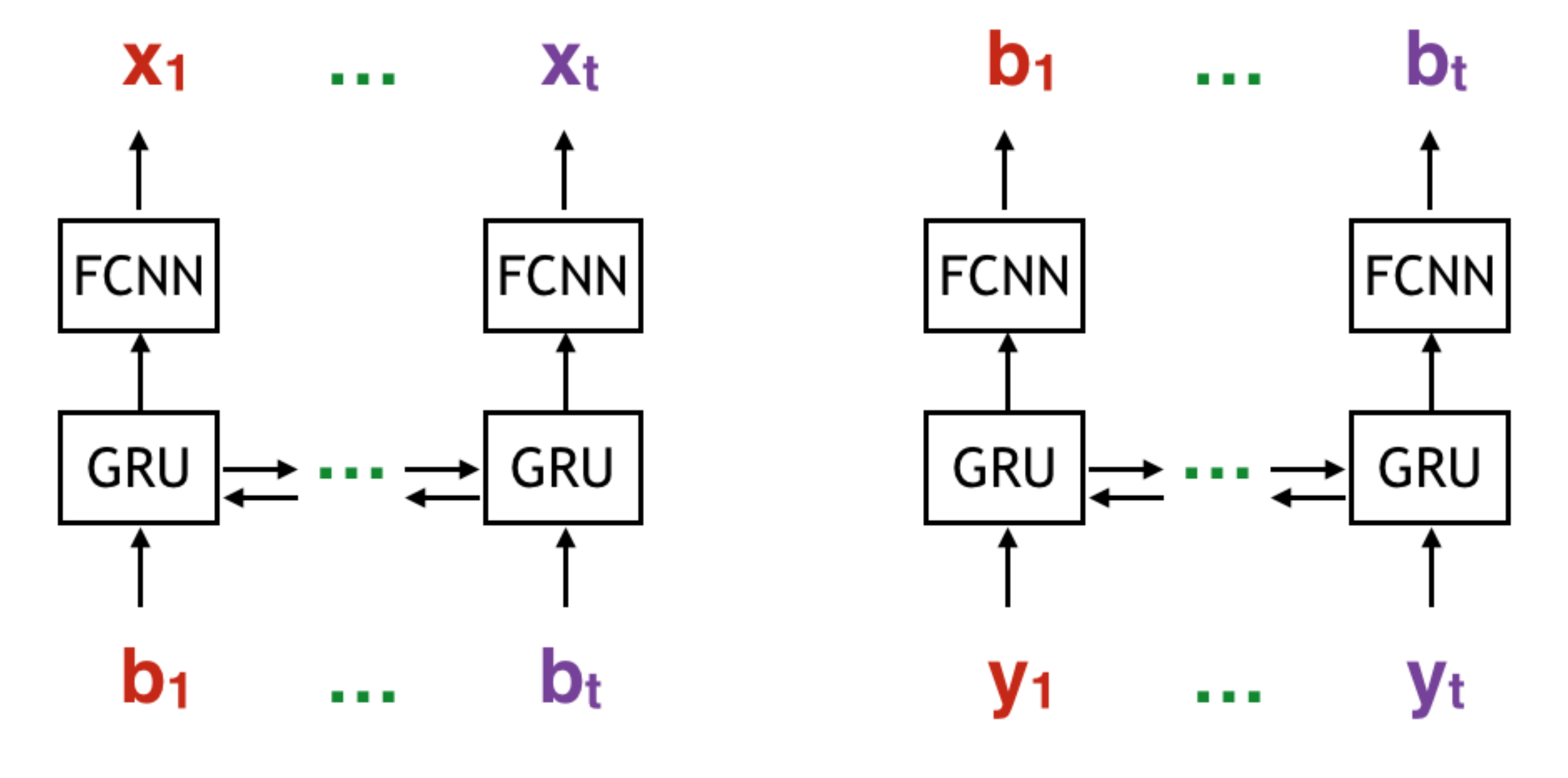}\ \ \ 
\begin{center}
\begin{tabular}{ |c|c| } 
 \hline
 Decoder layer & Output dimension \\ 
 \hline
 input & (K, 1/R) \\ 
 Bi-GRU (2 layers) & (K, 100) \\ 
 FCNN (sigmoid) & (K,1)\\
 \hline
\end{tabular}
\end{center}
\begin{center}
\begin{tabular}{ |c|c| } 
 \hline
 Encoder layer & Output dimension \\ 
 \hline
 input & (K, 1) \\ 
 Bi-GRU (2 layers) & (K, 25) \\ 
 FCNN (linear) & (K,1/R)\\
 \hline
\end{tabular}
\end{center}

\caption{RNN-based Channel AE encoder (top left), decoder (top right), and network structures (bottom).}\label{earn_structure}
\end{figure}

Applying the network architecture guidelines and the training methodology improvements thus far proposed, we design a neural code with Bi-GRU for both encoder and decoder, as shown in Figure~\ref{earn_structure}. The hyperparameters are shown in Figure~\ref{earn_hyper}. 

\begin{figure}[!ht] 
\centering
{\small
 \begin{tabular}{|c c |} 
 \hline
 Encoder & 2-layer Bi-GRU w/ 25 units   \\ 
 Decoder & 2-layer Bi-GRU w/ 100 units  \\ 
 Power constraint  & bit-wise normalization \\
 Batch size & 1000\\
 Learning rate & 0.001, $\times$ 1/10 when saturate \\
 Num epoch & 250 \\
 Block length & 100 \\
 Batch per epoch & 100\\
 Optimizer & Adam \\
 Loss  & Binary Cross Entropy (BCE)\\
 Min Dist Regularizer  & 0.0 \\
 Train SNR at rate 1/2 & mixture of 0 to 8dB \\
 Train SNR at rate 1/3 & mixture of -1 to 2dB \\
 Train SNR at rate 1/4 & mixture of -2 to 2dB \\  
 Train method& train ENC once DEC 5 times \\
 Min Distance Regularizer  &  0.001 ($s = 10$) \\ 
 \hline
\end{tabular}
}
\centering
\caption{RNN-based Channel AE hyperparameters.}\label{earn_hyper}
\end{figure}

\begin{figure*}[!ht] 
\centering
\includegraphics[width=0.90\textwidth]{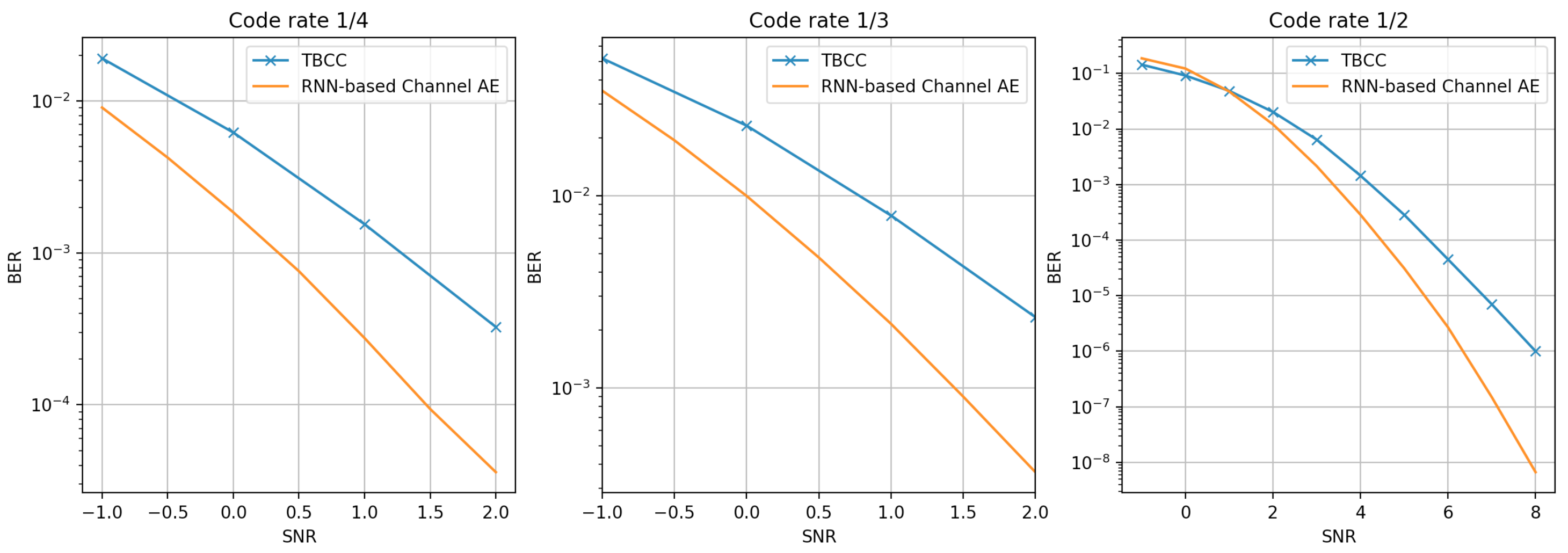}\ \ \ 
\vspace{-1em}
\caption{Comparing RNN-based Channel AE with convolutional code on rate 1/4 (left), 1/3 (middle), 1/2 (right). Block length  = 100.}\label{bl_awgn}
\end{figure*}

Tail-biting Convolutional Code (TBCC) has proven to be the state of the art under a short block length regime \cite{rachinger2015comparison}, \cite{liva1, liva2, liva3}.
We compare the performance of TBCC with RNN-based channel code on block code settings.
The BER performance in the AWGN channel under various code rates is shown in Figure~\ref{bl_awgn}. 
The TBCC BER curve is generated by convolutional code with constraint length up to $m=7$ by the best generator function from Figure~\ref{ll_conv}, with traceback length equals $5(m+1)$; we measure the BERs for all TBCCs in Figure~\ref{ll_conv} empirically using CommPy simulator~\cite{commpy} and plot the best performing one. 
Figure~\ref{bl_awgn} shows that RNN-based Channel AE outperforms all convolutional codes up to constraint length 7. RNN-based Channel AE empirically demonstrates the advantage of jointly optimizing encoder and decoder over the AWGN channel. 

Note that the RNN-based Channel AE code is continuous; the performance gain is from both coding gain and high-order modulation gain, as shown in Figure~\ref{bl_awgn} right; the performance gap at a higher SNR is larger. 
Binarized code leads to a fairer comparison; however, binarizing with the sign function is not differentiable. Bypassing non-differentiability using a straight-through estimator (STE)~\cite{bnn2016} degrades performance in channel coding~\cite{turboae2019}.
Binarizing code and comparing to high order modulation are deferred to future research.

\subsection{Performance of RNN-Based Channel AE: Non-AWGN Setting}

We test the robustness and adaptivity of RNN-based Channel AE on three families of channels:
\begin{enumerate}
\item AWGN channel: $y = x + z$, where $z \sim N(0, \sigma^2)$.

\item Additive T-distribution Noise (ATN) channel: $y = x + z$, where $x \sim T(v, \sigma^2)$, for $v = 3, 5$. 
This heavy-tailed noise distribution models bursty interference. 

\item Radar channel: $y = x + w+ z$, where $z \sim N(0, \sigma_1^2)$ and $w \sim N(0, \sigma_2^2)$, w.p. $ p=0.05$. (Assume $\sigma_1 \ll \sigma_2$). This noise model appears when there is bursty interference, for example, when Radar interferes with LTE \cite{sanders2014effects,safavi2015impact}.
\end{enumerate}

\noindent\textbf{Robustness}\\
Robustness refers to the property that when RNN-based Channel AE is trained for the AWGN channel, the test performance with no re-training on a different channel (ATN and Radar) should not degrade significantly. Most existing codes are designed under AWGN since it has a clean mathematical abstraction, and AWGN is the worst case noise under a given power constraint~\cite{shannon2001mathematical}. When both the encoder and decoder are unaware of the non-AWGN channel statistics, the BER performance degrades. 
Robustness ensures both the encoder and decoder perform well under channel mismatch, which is a typical use case for the low-latency scheme when channel estimation and corresponding receiver calibration are not accurate~\cite{richardson2008modern}.

\noindent\textbf{Adaptivity}\\
Adaptivity allows an RNN-based Channel AE to learn a decoding algorithm from sufficient data (includes retraining) even when the channel statistics do not admit a simple mathematical characterization~\cite{kim2018communication}. We train RNN-based Channel AE under ATN and Radar channels with the same hyperparameters shown in Figure \ref{earn_hyper} and the same amount of training data to ensure that the RNN-based Channel AE converges. With both learnable encoder and decoder, two cases of adaptivity are tested. First is decoder adaptivity, where the encoder is fixed and the decoder can be further trained. Second is the full adaptivity of both encoder and decoder. In our findings, encoder adaptivity shows no further advantage and is thus omitted. 

\begin{figure*}[!ht] 
\centering
\includegraphics[width=0.9\textwidth]{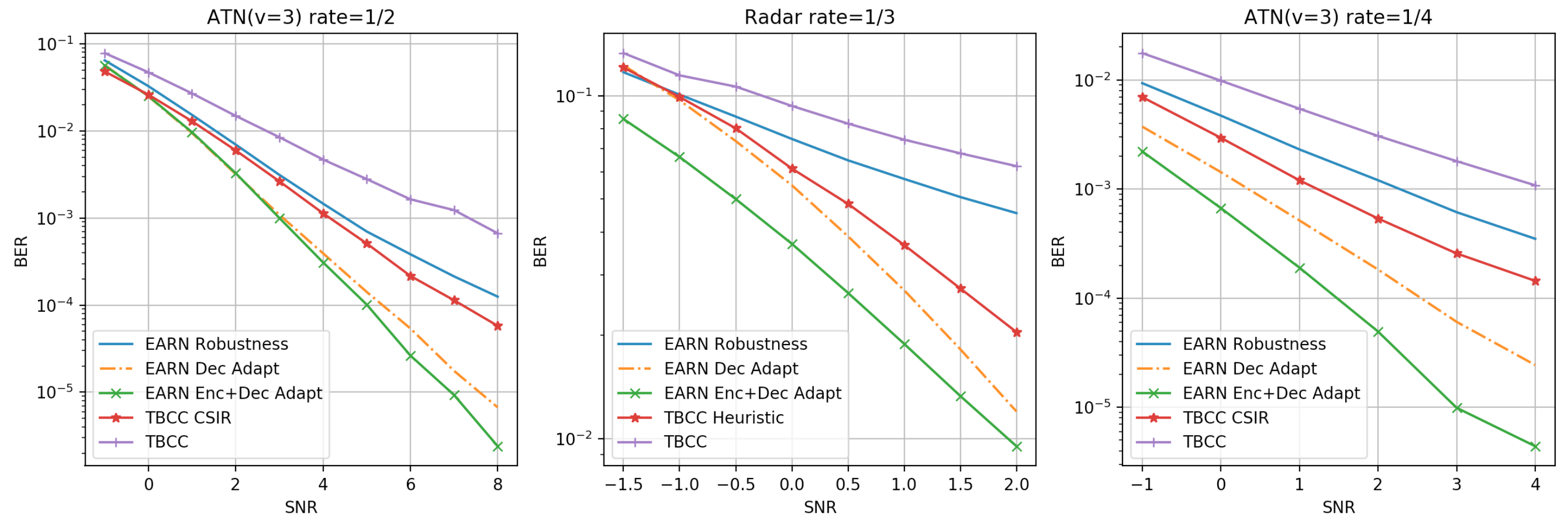}\ \ \ 
\caption{RNN-based Channel AE vs TBCC, ATN($\nu=3$) at rate 1/2 (left), Radar ($\sigma_2^2 = 5.0$) at rate 1/3 (middle), ATN($\nu=3$) at rate 1/4 (right). TBCC is m=2, with $g_{11}=5$, $g_{12}=7$ with feedback $7$}\label{bl_nonawgn}
\end{figure*}

We now evaluate RNN-based Channel AE for robustness and adaptivity on ATN and Radar channels. The BER performance is shown in Figure~\ref{bl_nonawgn}. Note that under non-AWGN channels, RNN-based Channel AE trained on the AWGN channel outperforms the convolutional code with the Viterbi decoder. It also shows more robust decoding ability for channel mismatching compared to the best convolutional code. As shown in Figure~\ref{bl_nonawgn}, RNN-based Channel AE with decoder-only adaptivity improves over the RNN-based Channel AE robust decoder, while RNN-based Channel AE with full adaptivity with both trainable encoder and decoder shows the best performance. 

The fully adapted RNN-based Channel AE outperforms the convolutional code even with CSIR, which utilizes the log-likelihood of T-distribution noise. Thus, designing jointly by utilizing the encoder and decoder further optimizes the code under given channels. Even when the underlying mathematical model is far from a cleaner abstraction, RNN-based Channel AE can learn the underlying functional code via self-supervised back-propagation.

RNN-based Channel AE is the first neural code to our knowledge that outperforms existing canonical codes under the AWGN channel coding setting, which opens a new field of constructing efficient neural codes under canonical settings. Furthermore, RNN-based Channel AE can be applied to channels when, especially, closed form mathematical analysis cannot be performed. 


\section{Designing Low Latency Neural Codes: LEARN}\label{learn}

Designing codes for low latency constraints is challenging since many existing block codes require inevitably long block lengths. 
To address this challenge, we now propose a new RNN-based encoder and decoder architecture that satisfies a low latency constraint, which we call LEARN. 
While our encoder and decoder architectures are based on RNNs, we introduce a new block in the decoder architecture so that the decoder can satisfy the extreme low latency constraint.
We show that LEARN code is: (1) significantly more reliable than convolutional codes, which are state-of-the-art under extreme low latency constraints~\cite{rachinger2015comparison}, and (2) more robust and adaptive for various channels beyond AWGN channels. In the following, we first define latency and review the literature for the low latency setting.
 
\subsection{low latency Convolutional Code}
Formally, decoder structural delay $D$ is understood in the following setting: to send message $b_t$ at time $t$, the causal encoder sends code $x_t$, and the decoder has to decode $b_t$ as soon as it receives $y_{t+D}$. The decoder structural delay $D$ is the number of bits that the decoder can look ahead to decode. The convolutional code has $0$ encoder delay due to its causal encoder, and the decoder delay is controlled by the optimal Viterbi decoder~\cite{viterbi1967error} with a decoding window of length $w$, which uses only the next $w$ future branches in the trellis to compute the current output. For code rate $R = \frac{k}{n}$ convolutional code, the structural decoder delay is $D = k-1 + kw$~\cite{maiya2012low}. 
 {\color{black}We use code rates $R=1/2,1/3,1/4$ ($k=1$ with $n = 2,3,4$).}; the structural decoder delay is $D = w$. 
Convolutional code is state-of-the-art code under extreme low latency, where $D\leq 50$~\cite{rachinger2015comparison}. 

\begin{figure}[!ht] 
\tiny{
\centering
\begin{tabular}{| p{0.3cm} |p{0.3cm} |p{0.3cm} |p{0.3cm} |p{0.3cm}|p{0.3cm}|p{0.3cm}|p{0.3cm}| p{0.3cm} |p{0.3cm} |p{0.5cm} | }
 \hline
    & \multicolumn{2}{|c|}{R=1/2}   & \multicolumn{3}{|c|}{R=1/3}&   \multicolumn{4}{|c|}{R=1/4} &\\
 \hline
 m  & $g_{11}$ & $g_{12}$ & $g_{11}$ & $g_{12}$  & $g_{13}$  & $g_{11}$ & $g_{12}$  & $g_{13}$ & $g_{14}$ &Feedback \\
  \hline
 1 & 2 &3 & 1 & 3 & 3 & 1 & 1 & 3 & 3  & 3\\
 2 & 5 &7 & 5 & 7 & 7 & 5 & 7 & 7 & 7  & 7\\
 3 & 15 &17 & 13 & 15 & 17 & 13 & 15 & 15 & 17  & 17\\
 4 & 23 & 35 & 25 & 33 & 37 & 25 & 27 & 33 & 37 & 37  \\
 5 & 53 & 75 & 47 & 53 & 75 & 53 & 67 & 71 & 75 & 75  \\
 6 & 133 & 171 & 133 & 145 & 175 & 135 & 135 & 147 & 163 & 163  \\
 7 & 247 & 371 & 225 & 331 & 367 & 237 & 275 & 313 & 357 & 357  \\
 \hline
\end{tabular}
\includegraphics[width=0.15\textwidth]{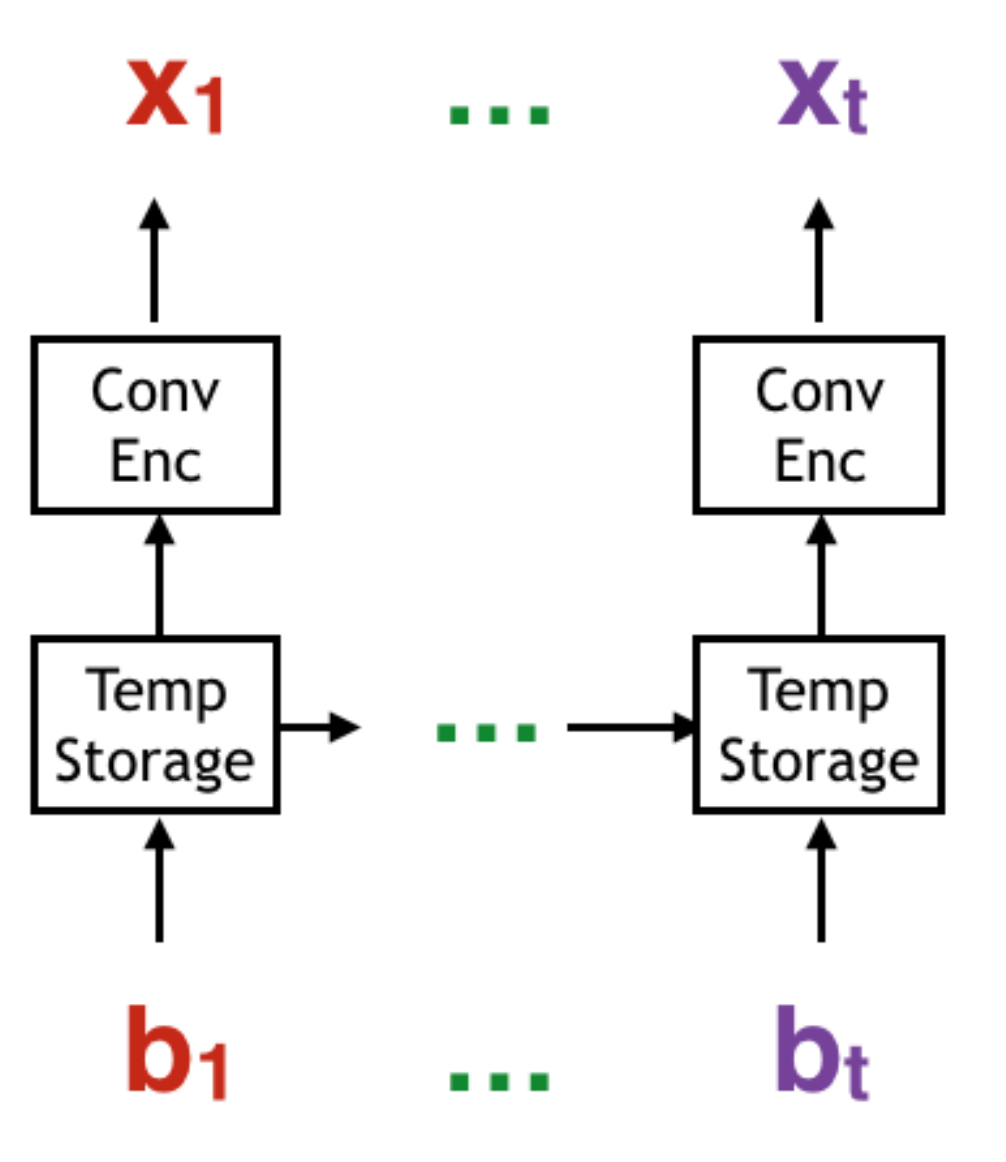}\ \ \ 
\caption{Convolutional code generator matrix in octal representation (top)~\cite{rachinger2015comparison} and encoder (bottom). 
}
\label{ll_conv}
}
\end{figure}

In this paper, we confine our scope to investigating extreme low latency with no encoder delay under low structural decoder delay $D = 1$ to $D = 10$ with code rates 1/2, 1/3, and 1/4. 
The benchmark we are using is convolutional codes with variable memory length. Under an unbounded block length setting, longer memory improves performance; however, under a low latency constraint, longer memory may not necessarily mean better performance since the decoding window is short ~\cite{rachinger2015comparison}. Hence, we test for all memory lengths under 7 to get the state-of-the-art performance of the Recursive Systematic Convolutional (RSC) code,  whose generating functions are shown in Figure \ref{ll_conv} (top), with the convolutional code encoder shown in Figure \ref{ll_conv} (bottom). The decoder is Viterbi with a decoding window $w = D$.  

\subsection{LEARN Network Architecture}
\begin{figure}[!ht] 
\centering
\includegraphics[width=0.15\textwidth]{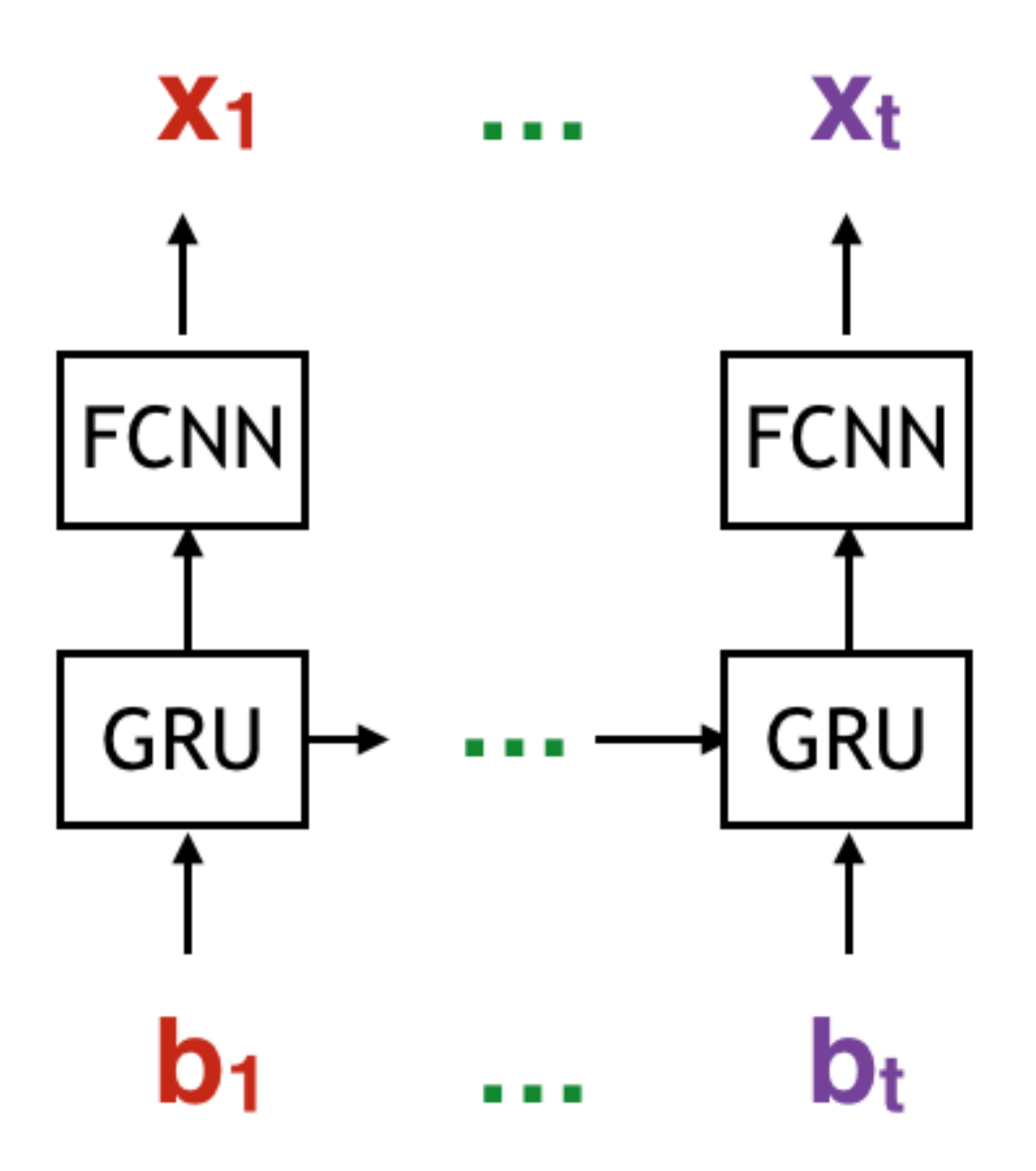}\ \ \ 
\includegraphics[width=0.27\textwidth]{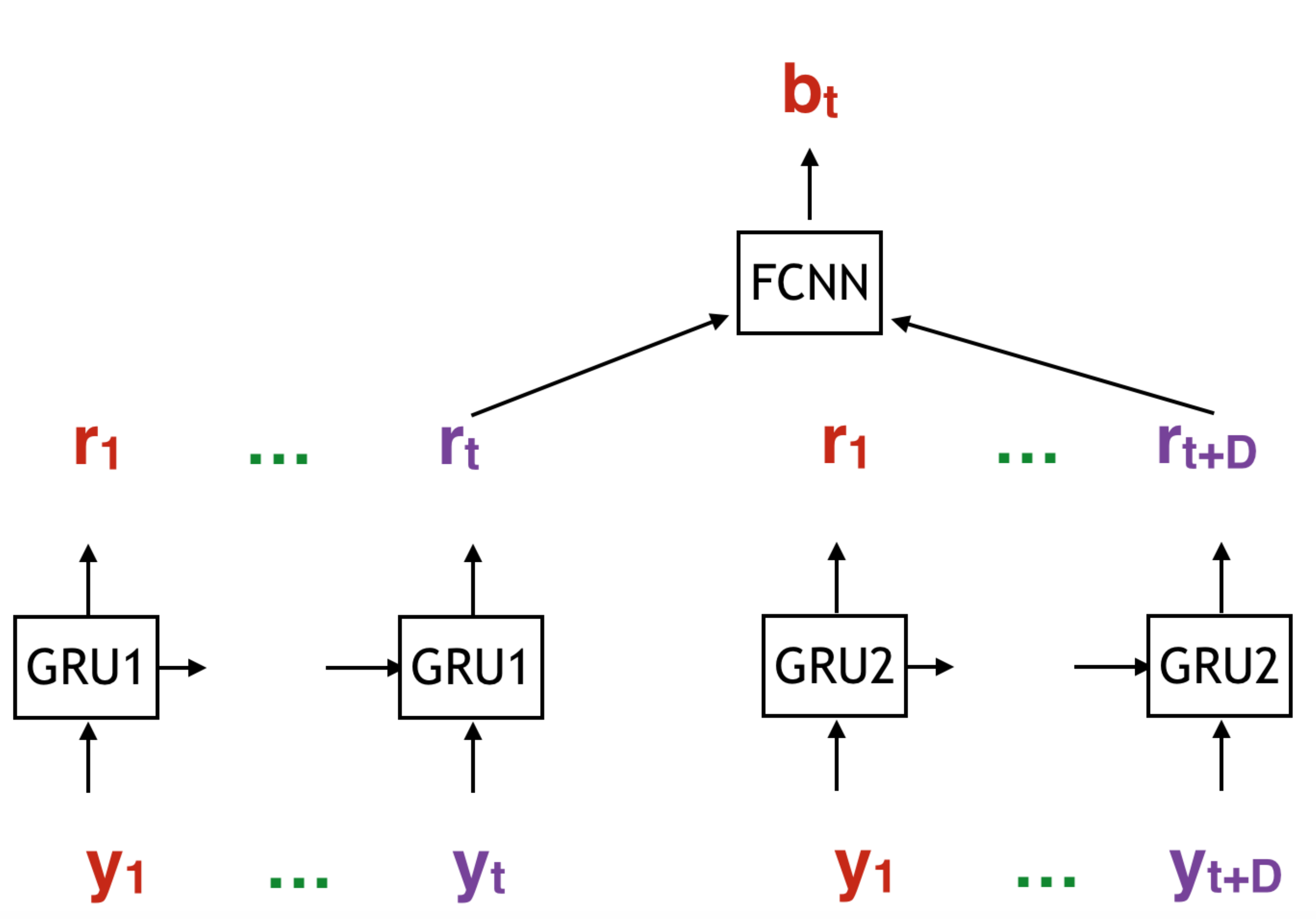}\ \ \ 
\begin{center}
\begin{tabular}{ |c|c| } 
 \hline
 Decoder layer & Output dimension \\ 
 \hline
 Input & (K, 1/R) \\ 
 GRU1 (2 layers) & (K, 100) \\ 
 GRU2 (2 layers) & (K, 100) \\ 
 FCNN (sigmoid) & (K,1)\\
 \hline
\end{tabular}
\end{center}
\begin{center}
\begin{tabular}{ |c|c| } 
 \hline
 Encoder layer & Output dimension \\ 
 \hline
 Input & (K, 1) \\ 
 GRU (2 layers) & (K, 25) \\ 
 FCNN (linear) & (K,1/R)\\
 \hline
\end{tabular}
\end{center}

\vspace{-0.5em}
\caption{LEARN encoder (top left), LEARN decoder (top right), and network structures (bottom).}\label{fig:learn}
\vspace{-0.5em}
\end{figure}

Using the network design proposed in the previous section, we propose a novel RNN-based neural network architecture for LEARN (both encoder and decoder) that satisfies the low latency constraint. Our proposed LEARN encoder is illustrated in Figure~\ref{fig:learn} (top left). The causal neural encoder is a causal RNN with two layers of GRU added to a Fully Connected Neural Network (FCNN). The neural structure ensures that the optimal temporal storage can be learned and extended to a non-linear regime. The power constraint module is bit-wise normalization, as described in the previous section. 

Applying the Bi-RNN decoder for low latency code requires the computation of lookahead instances for each received information bit, which is computationally expensive in both time and memory. To improve efficiency, the LEARN decoder uses two GRU structures instead of Bi-RNN structures. It has two GRUs: one GRU runs till the current time slot, and another GRU runs further for $D$ steps; the outputs of the two GRUs are then summarized by a FCNN. The LEARN decoder ensures that all the information bits satisfying the delay constraint can be utilized with the forward pass only. When decoding a received signal, each GRU needs to process only one step ahead, which results in decoding computation complexity $O(1)$. Viterbi and BCJR low latency decoders need to go through the trellis and backtrack to the desired position, which requires going forward one step and backward with delay constraints steps, resulting in $O(D)$ computation for decoding each bit. Although GRU has a large computational constant due to the complexity of the neural network, the computation time is expected to diminish~\cite{ovtcharov2015accelerating} with emerging AI chips. The hyperparameters of LEARN differ from block code settings. To summarize the differences: (1) the encoder and decoder use GRU instead of Bi-GRU, (2) the number of training epochs is reduced to 120, and (3) we do not use a partial minimum distance regularizer. 

\subsection{Performance of LEARN: AWGN Setting}
\begin{figure*}[!ht] 
\centering
\includegraphics[width=0.80\textwidth]{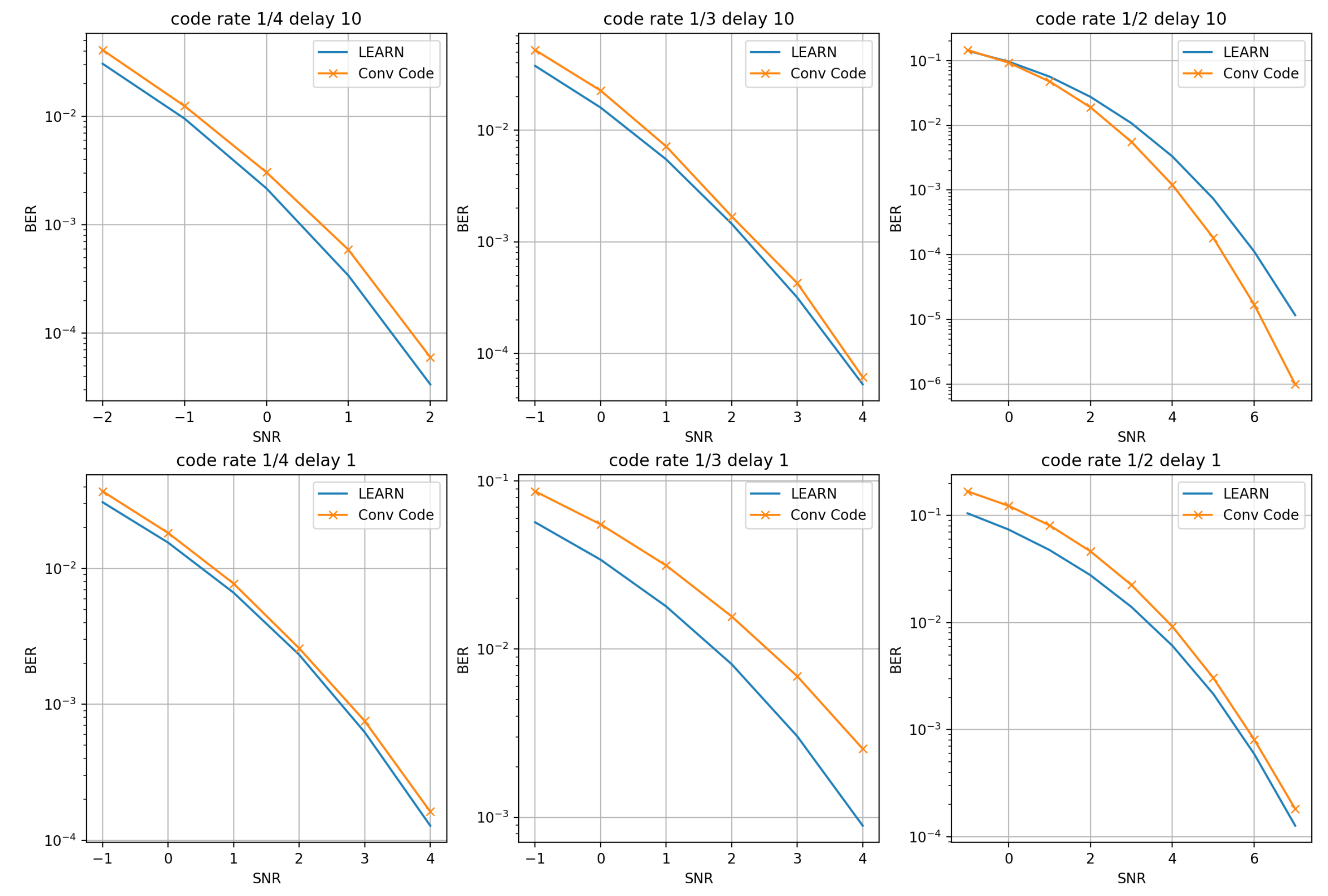}\ \ \ 
\vspace{-1em}
\caption{BER curves comparing low latency convolutional code vs LEARN under the AWGN channel with rate 1/4, 1/3, and 1/2.}\label{ll_gru}
\vspace{-1em}
\end{figure*}

Figure~\ref{ll_gru} shows the BER of a LEARN code and state-of-the-art RSC codes with varying memory lengths as illustrated in Figure~\ref{ll_conv} (top) for rates 1/2, 1/3, and 1/4 as a function of SNR under latency constraints $D=1$ and $D=10$. 
As the figure shows, for rates 1/3 and 1/4 under the AWGN channel, LEARN code under extreme delay ($D=1$ to $D=10$) shows better performance in Bit Error Rate (BER) compared to the state-of-the-art RSC codes from varying memory lengths in Figure \ref{ll_conv} (top). LEARN outperforms all RSC codes listed in Figure \ref{ll_conv} (top) with $D \leq 10$ with code rates 1/3 and 1/4, demonstrating 
a promising application of neural code under the low latency constraint. 

For higher code rates (such as $R =\frac{1}{2}$ and $D \geq 5$), LEARN shows comparable performance to convolutional codes but degrades at a high SNR. We expect further improvements can be made via improved structure design and hyperparameter optimization, 
 especially at higher rates.

\subsection{Robustness and Adaptivity}

\begin{figure*}[!ht] 
\centering
\includegraphics[width=0.9\textwidth]{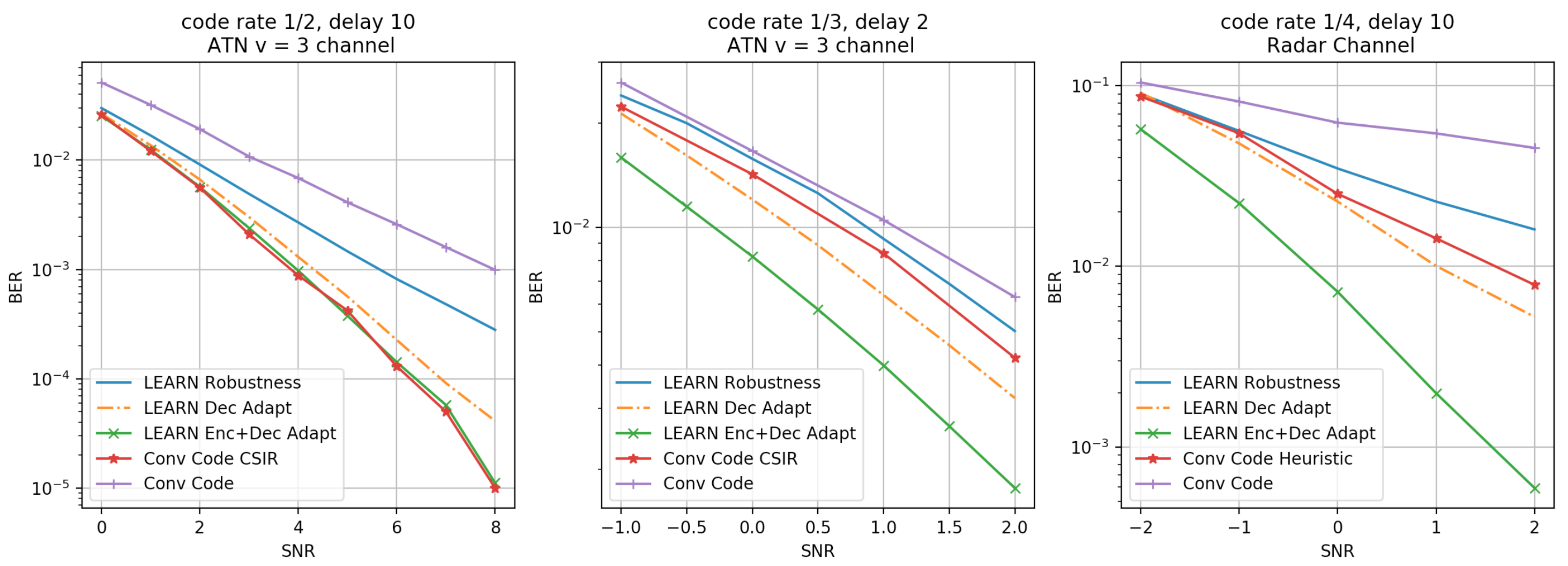} \ \ \
\caption{LEARN robustness and adaptivity in ATN ($\nu=3$, $R=1/2$, $D=10$)(left), ATN ($\nu=3$, $R=1/3$, $D=2$)(middle), and Radar ($p=0.05$, $\sigma_2 = 5.0$, $R=1/4$, $D=10$) (right) channels. }\label{LL_adapt}
\end{figure*}

The performance of LEARN with reference to robustness and adaptivity is shown in Figure~\ref{LL_adapt} for three different settings: (1) delay $D=10$, code rate $R = 1/2$, with ATN ($\nu = 3$) channel; (2) delay $D=2$, code rate $R = 1/3$, with ATN ($\nu = 3$) channel; and (3) delay $D=10$, code rate $R = 1/2$, with the Radar ($p=0.05$, $\sigma_2 = 5.0$) channel. As shown in Figure~\ref{LL_adapt} (left), with ATN ($\nu = 3$), which has a heavy-tail noise, LEARN with robustness outperforms convolutional codes. 
An improved adaptivity is achieved when both the encoder and decoder are trained, compared to when only decoder is trained. 
By exploring a larger space of codes, neural designed coding schemes can match canonical convolutional codes with Channel State Information at Receiver (CSIR) at a low code rate ($R=1/2$) and outperform convolutional codes with CSIR at a high code rate ($R=1/3$). 

As for Figure~\ref{LL_adapt} (middle) ATN ($\nu = 3$) channel with code rate $R=1/3$ and Figure~\ref{LL_adapt} (right) Radar ($\sigma_2 = 5.0$) channel with code rate $R=1/4$, the same trend holds. Note that under the Radar channel, we apply the heuristic proposed in \cite{safavi2015impact}. We observe that LEARN with full adaptation gives an order-of-magnitude improvement in reliability over the convolutional code heuristic \cite{safavi2015impact}.  This experiment shows that by jointly designing both encoder and decoder, LEARN can adapt to a broad family of channels. LEARN offers an end-to-end low latency coding design method that can be applied to any statistical channels and ensure good performance. 

\vspace{-1em}

\section{Interpretability of Deep Models}\label{interpret}
The promising performance of LEARN and the RNN-based Channel AE leads to a question: how do we interpret what the encoder and decoder have learned? Answering this can inspire future research as well as help us find caveats and limitations. In this section, we present interpretation analysis via local perturbation for LEARN and RNN-based Channel AE encoders and decoders. 

\subsection{Encoder Interpretability }
The \textit{significant recurrent length} of RNNs is a recurrent capacity indicator that helps to interpret neural encoder and decoders. The RNN encoder's significant recurrent length is defined, at time $t$, as how long a sequence the input $u_t$ can impact as RNN operates recurrently.  Assume two input sequences, $u_1 = {u_{1,1}, ..., u_{1, t}, ... u_{1, T}}$ and $u_2 = {u_{2,1}, ..., u_{2, t}, ... u_{2,T}}$, where only $u_{1,t}$ and $u_{2, t}$ differ. Taking a batch of $u_1$ and $u_2$ as input for the RNN encoder, we compare the output absolute difference along the whole block between $x_{1} = f(u_1)$ and $x_{2} = f(u_2)$ to investigate how long the input flip at time $t$ can affect. 
To investigate LEARN's RNN encoder, we flip only the first bit (position 0) of $u_1$ and $u_2$. The code position refers to the block bit positions, and the y-axis shows the averaged difference between two different sequences. Figure \ref{dp_len} (top left) shows that for an extremely short delay $D=1$, the encoder's significant recurrent length is short. The effect of the current bit diminishes after 2 bits. As the delay constraint increases, the encoder's significant recurrent length increases, accordingly. The LEARN encoder learns to encode locally to optimize under the low latency constraint.

For RNN-based Channel AE with a Bi-RNN encoder, the block length is 100, and the flip is applied at the middle 50th bit position. Figure \ref{dp_len} (top right) shows the encoder trained under the AWGN and ATN channels. The encoder trained on ATN shows a longer significant dependency length.
ATN is a heavy-tail noise ditribution. To alleviate the effect of extreme values, increasing the dependency of an encoder helps. 
Note that even the longest significant recurrent length is only backward 10 steps and forward 16 steps; thus, the GRU encoder actually did not learn a very long recurrent dependency. AWGN capacity-achieving codes have some inbuilt mechanisms to improve long-term dependency. For example, the Turbo encoder uses an interleaver to improve the long-term dependency~\cite{berrou1993near}. Improving the significant recurrent length length via a more learnable structure design is an interesting future research direction. 

\subsection{Decoder Interpretability}
\begin{figure}[!ht] 
\centering
\includegraphics[width=0.5\textwidth]{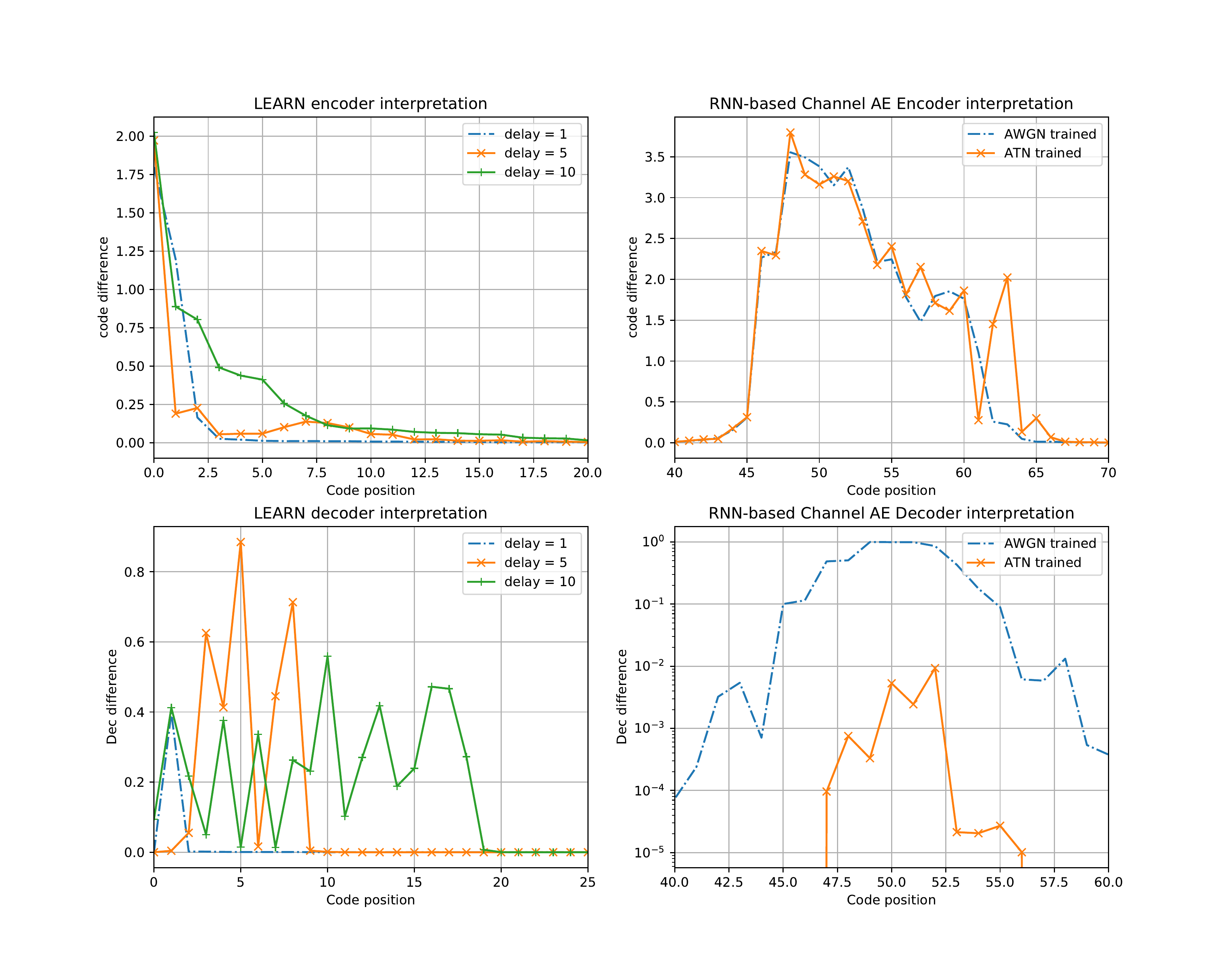}\ \ \
\vspace{-2em}
\caption{Encoder interpretation for LEARN (top left) and RNN-based Channel AE (top right). Decoder interpretation for LEARN (bottom left) and RNN-based Channel AE (bottom right).}\label{dp_len}
\end{figure}

The decoder's significant recurrent length can illustrate how it copes with different constraints and channels. Assume two noiseless coded sequences, $y_1 = {y_{1,1}, ..., y_{1, t}, ... y_{1, T}}$ and $y_2 = {y_{2,1}, ..., y_{2, t}, ... y_{2,T}}$, $y_1$, and $y_2$ equals $y_1$ other than at time $t$, where $y_{1, t} = y_{2,t} + p$, where $p$ is the large deterministic pulse noise; $p=5.0$ for our experiment. We compare the output absolute difference along the whole block between $\hat u_{1} = g(y_1)$ and $u_{2} = g(y_2)$ to investigate how long the pulse noise can affect. 
For the LEARN decoder, we inject pulse noise at the starting position. Figure \ref{dp_len} (bottom left) shows that for all delay cases, the noise most significantly affected the position equal to the delay constraint. This shows that the LEARN decoder learns to coordinate with the causal LEARN encoder. Since $D=1$, the maximized decoder difference along the block is at position 1; when $D=10$, the maximized decoder difference along the block is at position 10. Other code bits have a less significant but non-zero decoder difference. 

The LEARN decoder's significant recurrent length implies that it not only learns to utilize the causal encoder's immediate output, but it also utilizes outputs in other time slots to help decoding. Note that the maximized significant recurrent length is approximately twice the delay constraint; after less than approximately $2D$, the impact diminishes. The LEARN decoder learns to decode locally to optimize under different low latency constraints.
For RNN-based Channel AE with the Bi-RNN encoder, the perturbation is applied at the middle 50th position, still with block length 100. Figure \ref{dp_len} (bottom right) shows the decoder trained under the AWGN and ATN channels. The AWGN-trained decoder is more sensitive to pulse noise with extreme values {\ol than the ATN-trained decoder}. {\ol By reducing the sensitivity for extreme noise, the ATN-trained decoder learns to alleviate the effect of non-Gaussian noise.}
The RNN-based Channel AE decoder learns to optimize  under different channel settings.

\section{Conclusion}\label{conclusion}
In this paper, we demonstrated the power of neural network-based architectures to achieve state-of-the-art performance for joint encoder and decoder design. We showed that our learned codes significantly outperform convolutional codes in short to medium block lengths. However, in order to outperform state-of-the-art codes such as Turbo or LDPC codes, we require additional mechanisms such as interleaving to introduce long-term dependencies. This promises to be a fruitful direction for future exploration. 

In the low-latency regime, we achieved state-of-the-art performance with LEARN codes. Furthermore, LEARN codes beat existing codes by an order of magnitude in reliability when there is a channel mismatch. Our present design is restricted to very low structural latency; however, with additional mechanisms for introducing longer term dependencies~\cite{sutskever2014sequence,jaderberg2015spatial}, we believe that it is possible to extend these designs to cover a larger range of delays. This is another interesting direction for future work. 
Finally, we have focused only on very short structural delay. Latency depends on other factors as well (e.g., computational complexity). Optimization of neural decoders to reduce other attributes of latency poses an interesting open problem. 



\section*{Acknowledgment}
This work was supported in part by NSF awards CCF-1651236, CIF-1908003, CIF-1703403, CNS-2002932, CNS-2002664, and Intel. We also want to thank reviewers for the helpful and supportive comments.

\ifCLASSOPTIONcaptionsoff
  \newpage
\fi

\begin{IEEEbiographynophoto}{Yihan Jiang}
Yihan Jiang is a Ph.D. candidate at the Electrical and Computer Engineering Department, University of Washington, Seattle at Washington. He received his M.S degree in 2014 in Electrical and Computer Engineering from UC San Diego and a B.S degree from Beijing Institute of Technology in 2012. His research interests are in the areas of channel coding, information theory, deep learning, and federated learning.
\end{IEEEbiographynophoto}
\begin{IEEEbiographynophoto}{Hyeji Kim}
Hyeji Kim is a researcher at Samsung AI Research Cambridge. Prior to joining Samsung AI Research in 2018, she was a postdoctoral research associate at the University of Illinois, Urbana-Champaign. She received her Ph.D. and M.S. degrees in Electrical Engineering from Stanford University in 2016 and 2013, respectively, and her B.S. degree with honors in Electrical Engineering from KAIST in 2011. She is a recipient of the Stanford Graduate Fellowship and participant in the Rising Stars in EECS Workshop, 2015. 
\end{IEEEbiographynophoto}
\begin{IEEEbiographynophoto}{Himanshu Asnani}
Himanshu Asnani is currently Reader (eq. to tenure-track Assistant Professor) in the School of Technology and Computer Science at the Tata Institute of Fundamental Research, Mumbai. He is also an Affiliate Assistant Professor in the Electrical and Computer Engineering Department at University of Washington, Seattle, where he held a Research Associate position previously. His current research interests include machine learning, causal inference and information and coding theory. He received his M.S. and Ph.D. in Electrical Engineering Department in 2011 and 2014 respectively from Stanford University where he was a Stanford Graduate Fellow. He received his B.Tech. from IIT Bombay in 2009. Following his graduation, Himanshu worked briefly as a System Architect in Ericsson Silicon Valley following which he was involved in a couple education startups. In the past, he has also held visiting faculty appointments at Stanford University and IIT Bombay. He was the recipient of 2014 Marconi Society Paul Baran Young Scholar Award, 2018 Amazon Catalyst Fellow Award, Best Paper Award at MobiHoc 2009, and  was the Finalist for the Student Paper Award in ISIT 2011. 
\end{IEEEbiographynophoto}
\begin{IEEEbiographynophoto}{Sreeram Kannan}
Sreeram Kannan is currently an Assistant Professor at the University of Washington, Seattle. He was a postdoctoral scholar at the University of California, Berkeley between 2012-2014 before which he received his Ph.D. in Electrical Engineering and M.S. in Mathematics from the University of Illinois, Urbana-Champaign. He is a recipient of the 2019 UW ECE Outstanding Teaching award, 2017 NSF Faculty Early CAREER award, the 2013 Van Valkenburg outstanding dissertation award from UIUC, a co-recipient of the 2010 Qualcomm Cognitive Radio Contest first prize, a recipient of 2010 Qualcomm (CTO) Roberto Padovani Outstanding Intern award, a recipient of the SVC Aiya medal from the Indian Institute of Science, 2008, and a co-recipient of Intel India Student Research Contest first prize, 2006. His research interests include the applications of information theory and learning to blockchains, computational biology and wireless networks.
\end{IEEEbiographynophoto}
\begin{IEEEbiographynophoto}{Sewoong Oh}
Sewoong Oh is an Associate Professor in the Paul G. Allen School of Computer Science $\&$ Engineering at the University of Washington. Prior to joining the University of Washington in 2019, he was an Assistant Professor in the department of Industrial and Enterprise Systems Engineering at University of Illinois, Urbana-Champaign since 2012. He received his Ph.D. from the department of Electrical Engineering at Stanford University in 2011 under the supervision of Andrea Montanari. Following his Ph.D., he worked as a postdoctoral researcher at the Laboratory for Information and Decision Systems (LIDS) at MIT under the supervision of Devavrat Shah. Sewoong's research interest is in theoretical machine learning. He was co-awarded the ACM SIGMETRICS Best Paper award in 2015, the NSF CAREER award in 2016, the ACM SIGMETRICS Rising Star award in 2017, and the GOOGLE Faculty Research Award in 2017 and 2020.
\end{IEEEbiographynophoto}
\begin{IEEEbiographynophoto}{Pramod Viswanath}
Pramod Viswanath is a Professor of Electrical and Computer Engineering at the University of Illinois, Urbana-Champaign. He received his Ph.D. degree in Electrical Engineering and Computer Science from the University of California, Berkeley in 2000. His current research interests  include blockchain technologies from a variety of angles: networking protocols, consensus algorithms, payment channels, distributed coded storage and incentive designs. He is a co-founder and CEO of Applied Protocol Research, a startup focused on developing core blockchain technologies. He has received the Eliahu Jury Award, the Bernard Friedman Prize, a NSF CAREER award, and the Best Paper Award at the Sigmetrics conference in 2015.  He is a co-author, with David Tse, of the text Fundamentals of Wireless Communication, which has been used in over 60 institutions worldwide. 

\end{IEEEbiographynophoto}

\newpage
\newpage
    \setcounter{page}{0}
    \pagenumbering{arabic}
    \setcounter{page}{1}

\appendices
\section*{Appendix}
\label{appendix}
\section{Recurrent Neural Networks : An Introduction}
\label{appendix:rnn_intro}
\begin{figure}[!ht] 
\centering
\includegraphics[width=0.50\textwidth]{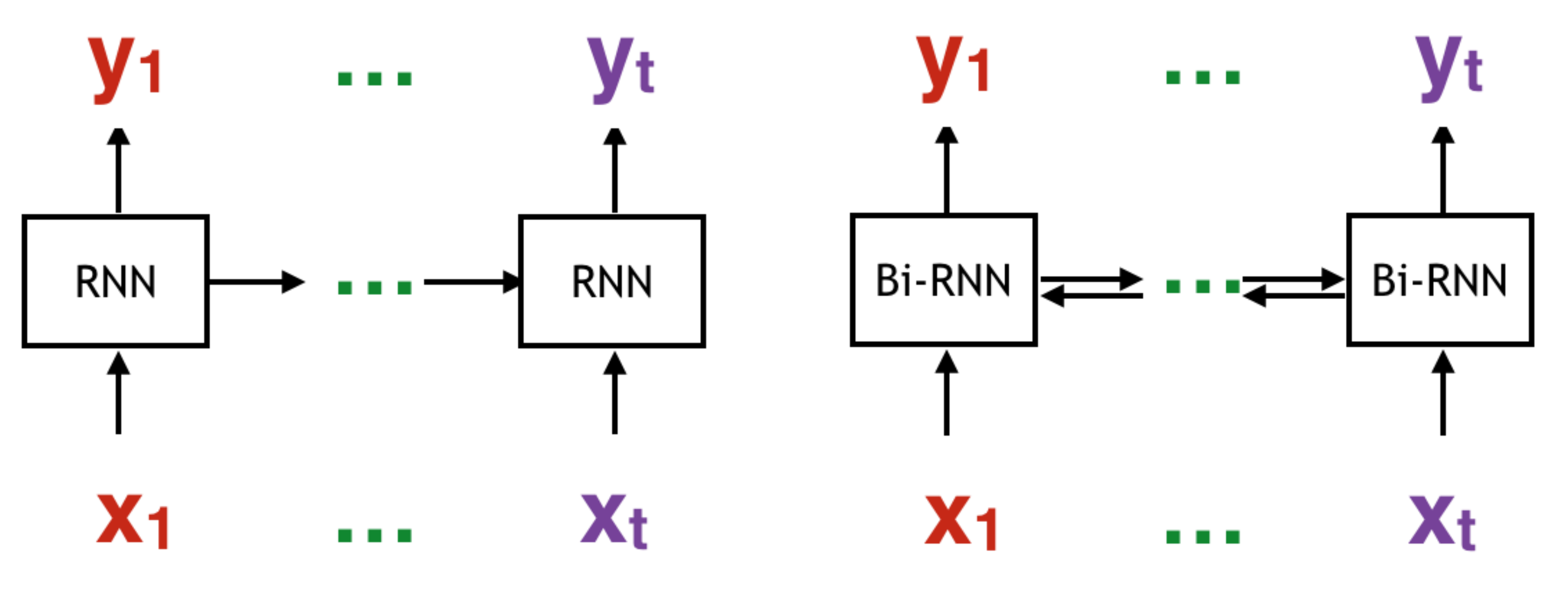}\ \ \
\includegraphics[width=0.47\textwidth]{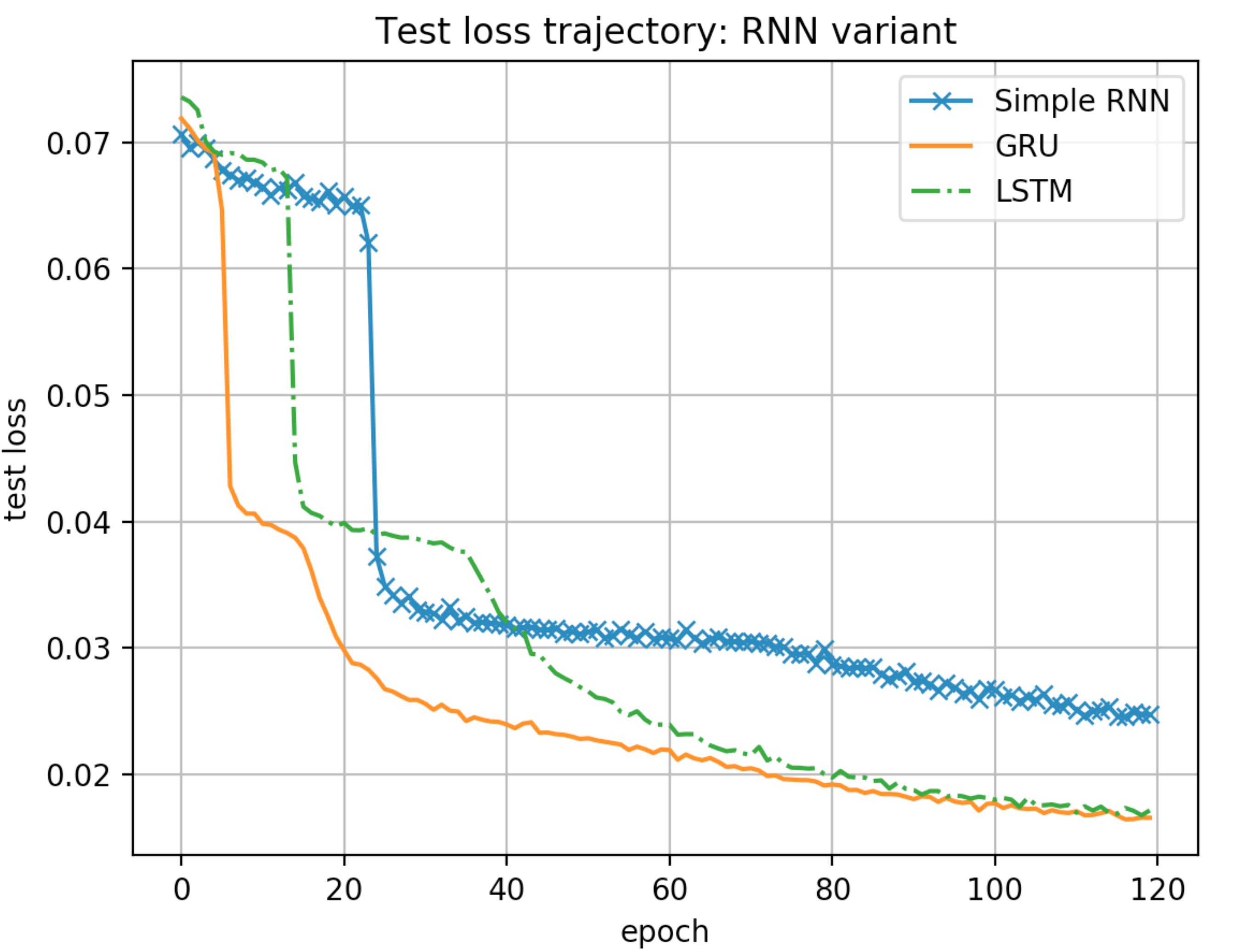}\ \ \
\caption{Basic RNN structure (top left), Bi-RNN (top right), and test trajectory for RNN variants selection (bottom)}\label{RNN_basic}
\end{figure}

As illustrated in Figure~\ref{RNN_basic} (top left), RNN is defined as a general function $f(.)$ such that $(y_t, h_t) = f(x_t, h_{t-1})$ at time $t$, where $x_t$ is the input, $y_t$ is the output, $h_t$ is the state sent to the next time slot and $h_{t-1}$ is the state from the last time slot. RNN can only emulate causal sequential functions. Indeed, it is known that an RNN can capture a general family of measurable functions from the input time sequence to the output time sequence \cite{hammer2000approximation}.  Illustrated in Figure~\ref{RNN_basic} (top right), Bidirectional RNN (Bi-RNN) combines one forward and backward RNN and can infer current state by evaluating through both past and future. Bi-RNN is defined as $(y_t, h^{f}_t, h^{b}_t) = f(x_t, h^{f}_{t-1}, h^{b}_{t+1})$, where $h^f_t$ and $h^{b}_t$ stands for the state at time $t$ for forward and backward RNNs~\cite{goodfellow2016deep}. 

RNN is a restricted structure which shares parameters between different time slots across the whole block. This makes it naturally generalizable to a longer block length. Moreover, RNNs can be considered as an over parameterized non-linear convolutional codes for both encoder and decoder, since a convolutional code encoder can be represented by a causal RNN and BCJR forward-backward algorithm can be emulated by a Bi-RNN \cite{kim2018communication}. 
There are several choices of parametric functions $f(.)$ for RNN, such as vanilla RNN, Gated Recurrent Unit (GRU), or Long Short Term Memory (LSTM). Vanilla RNN is known to be hard to train due to diminishing gradients. LSTM and GRU are the most widely used RNN variants which utilize gating schemes to alleviate the problem of diminishing gradients \cite{goodfellow2016deep}. We empirically compare Vanilla RNN with GRU and LSTM under test loss trajectory, which shows the test loss along with the training times. The test loss trajectory is the mean of 5 independent experiments. Figure \ref{RNN_basic} (bottom) depicts the test loss along training time, which shows that vanilla RNN has slower convergence and GRU converges fast, while GRU and LSTM have similar final test performances. Since GRU has less computational complexity, in this paper we use GRU as our primary network architecture \cite{chung2014empirical}. In this paper, we use the terms GRU and RNN interchangeably.  

\section{Power Constraint Module}
The power normalization enforces the output code to have a unit power by normalizing the output. In the following, we investigate three implementations on the power constraint layer that are differentiable. 
%
%
%
%
We let $b$, $\hat x$, $x$ denote the message bit sequence, output of the encoder, and the normalized codeword, respectively.   
%


\begin{enumerate}
\item[1.] {\em hard power constraint:} We use a hyperbolic tangent function in training ($x = tanh(\hat x)$) and threshold to -1 and +1 for testing, which only allows discrete coding schemes. 
\item[2.] {\em bit-wise normalization:} $E||X_i||_2^2 \leq 1, \forall i$. We have $x_i = \frac{\hat x_i - mean(x_i)}{std(x_i)} \ \forall i$. For a given coding bit position, the bit power is normalized.
\item[3.] {\em block-wise normalization}: $E||X||_2^2 \leq 1$,We have $x = \frac{\hat x- mean(x)}{std(x)} \  \forall i$ which allows us to reallocate power across the code block. 
\end{enumerate}

For bit-wise and block-wise normalizations, the behaviors in training and testing are different. During the training phase, bit-wise and block-wise normalizations normalize input mini-batch according to the training mini-batch statistics. During testing phase, the mean and standard deviation are precomputed by passing through many samples to ensure that the estimations of mean and standard deviation are accurate. 

Figure \ref{batch} left shows that block-wise normalization does offer better learning test trajectory, while bit-wise normalization shows slightly worse performance. Hard power constraint using $tanh$, due to saturating gradients, results in high test errors. To satisfy maximized power constraint, we use bit-wise normalization in this paper.

\section{Training Methodology}
\label{appendix:training}
In this section we give empirical evidence for the best effect of training techniques used in this paper. 
\subsection{Using Large Batch Sizes}
\begin{figure}[!ht] 
\centering
\includegraphics[width=0.45\textwidth]{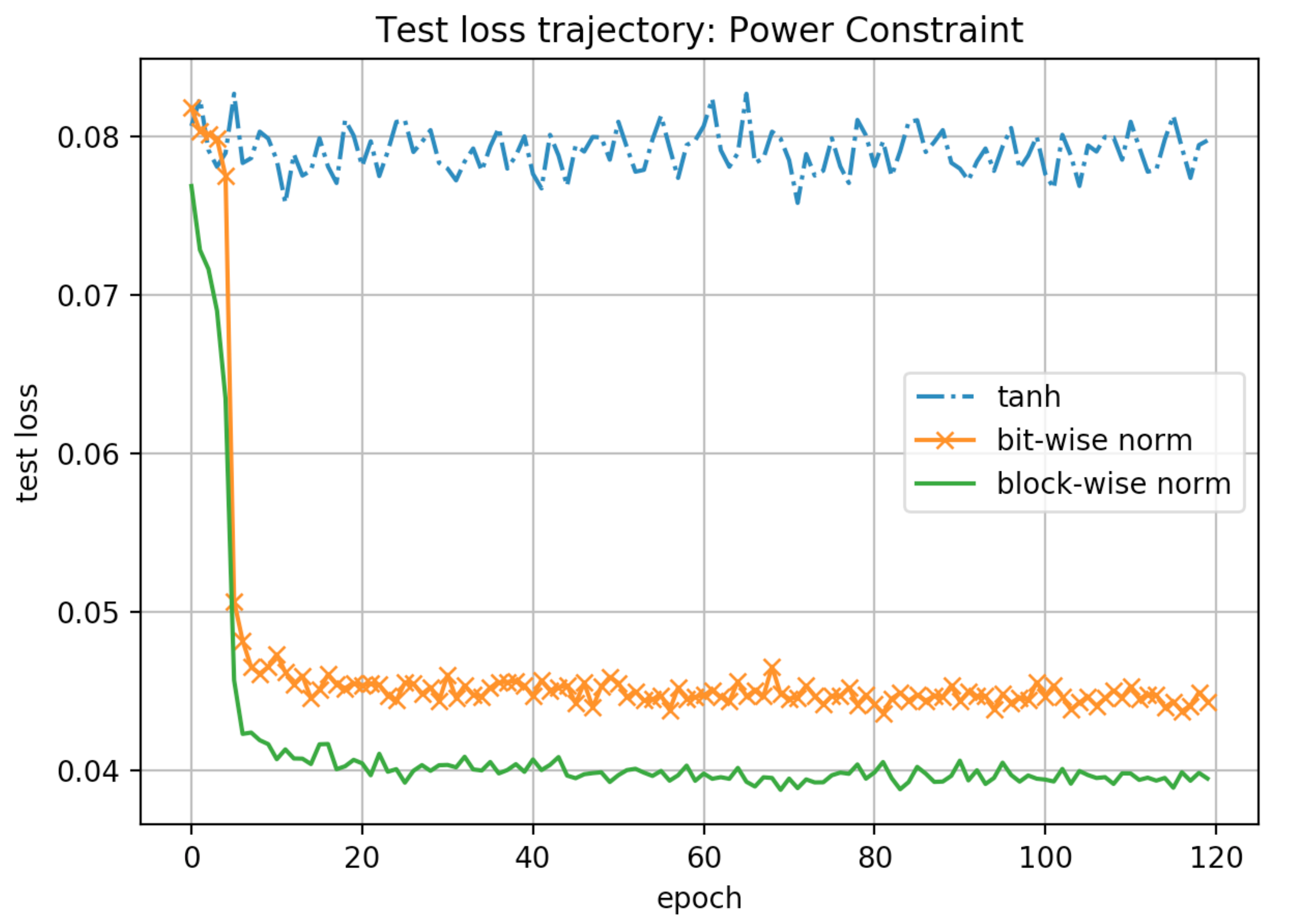}\ \ \ 
\includegraphics[width=0.45\textwidth]{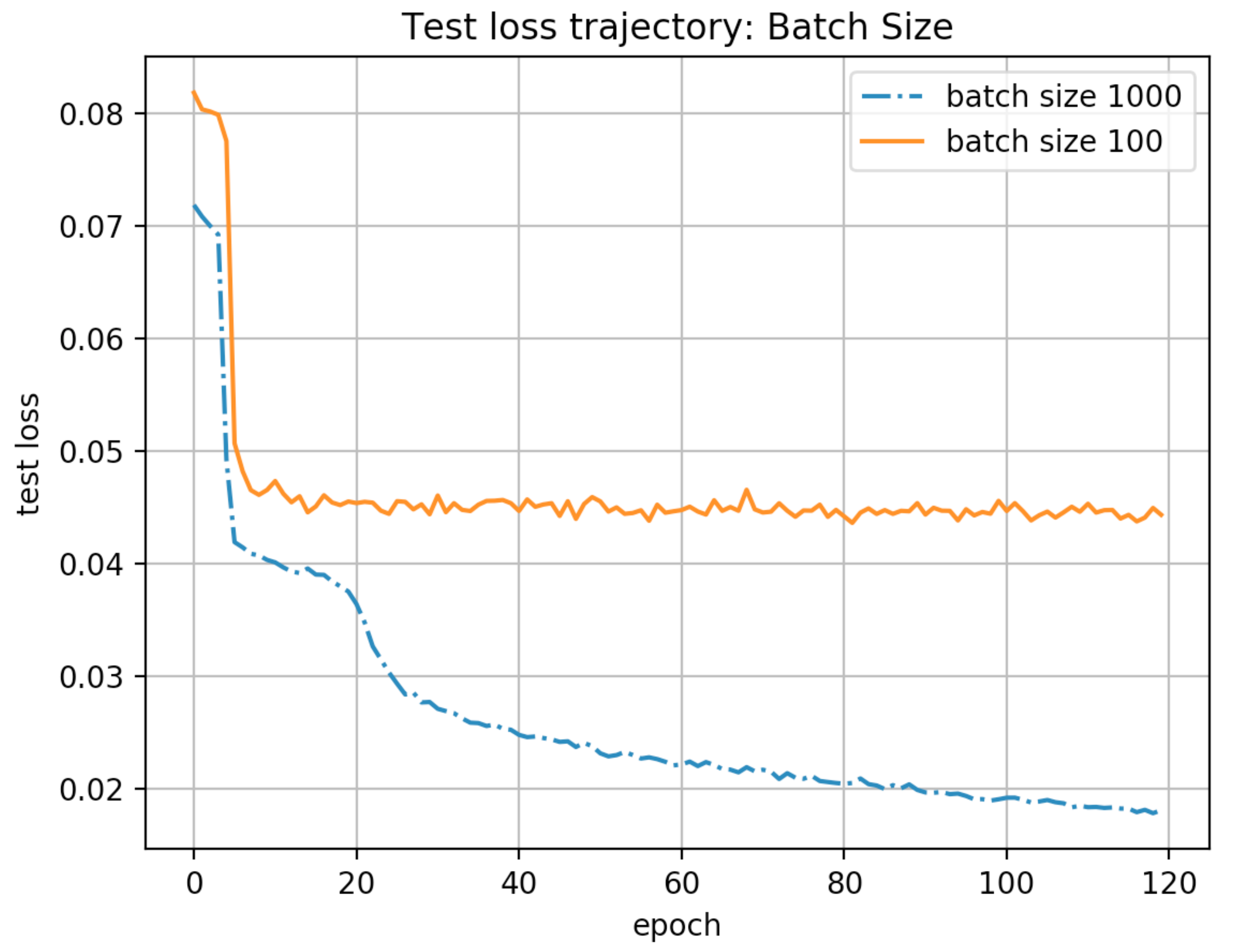}\ \ \ 
\includegraphics[width=0.45\textwidth]{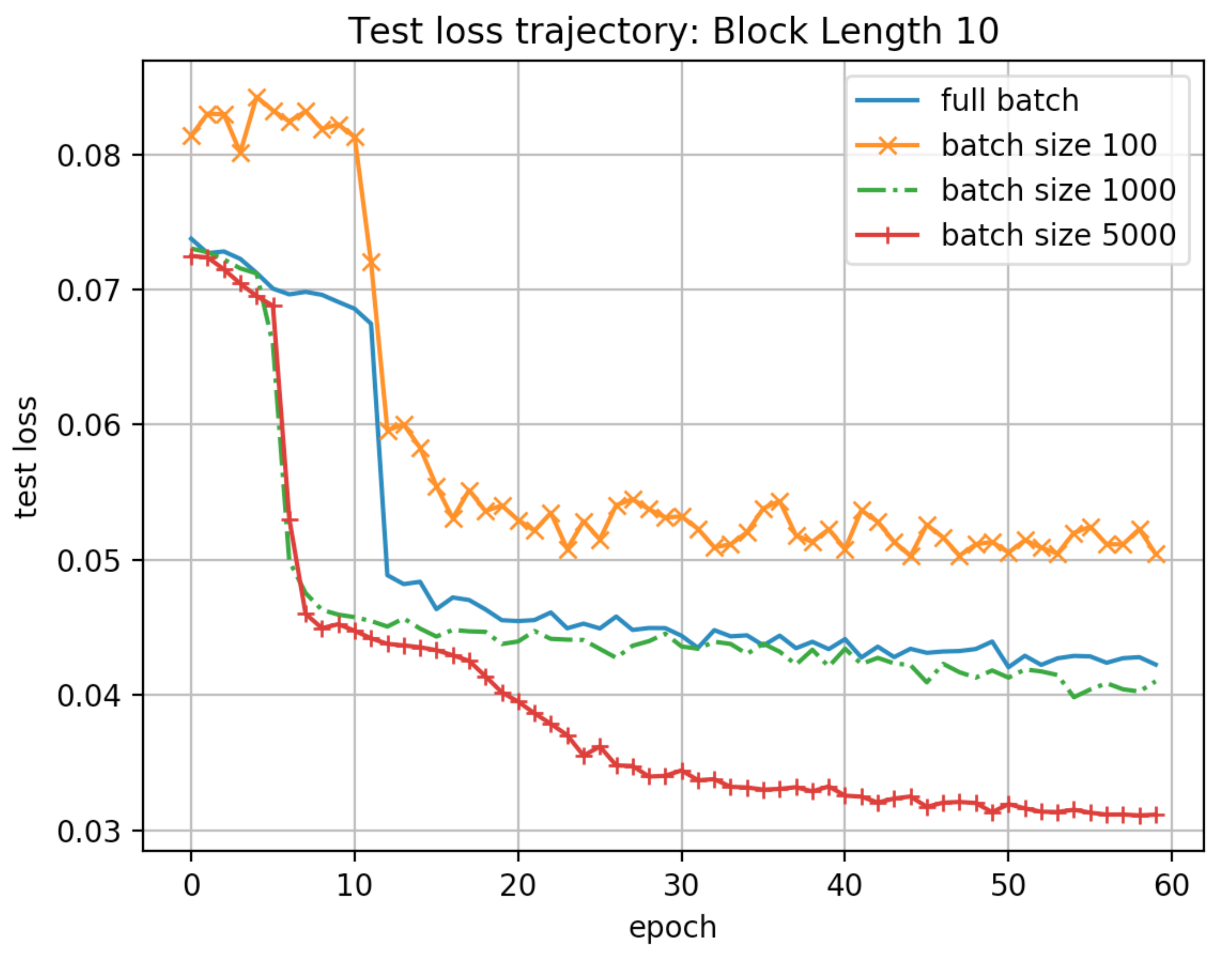}\ \ \ 
\caption{Test loss trajectory on different power constraint (top), different batch size with block length 100 (middle) and random batch with block length 10 (bottom)}\label{batch}
\end{figure}

Deep Learning models typically use mini-batching to improve generalization and reduce the memory usage. A small random mini-batch results in a better generalization \cite{hoffer2017train}\cite{smith2017don}, while a large batch size requires more training to generalize. However, Figure \ref{batch} (middle) shows that a larger batch size results in much better generalization for Channel AE, while small batch size tends to saturate in high test loss. A large batch size is required due to the following reasons:

(1) A large batch offers better power constraint statistics \cite{ioffe2015bn}. With a large batch size, the normalization of power constraint module offers a better estimation of mean and standard deviation, which makes the output of the encoder less noisy, thus the decoder can be better trained accordingly.

(2) A large batch size gives a better gradient. Since Channel AE is trained with a self-supervision, the improvement of Channel AE originates from error back propagation. As extreme error might result in wrong gradient direction, large batch size can alleviate this issue.

The randomness in the training mini-batch results in a better generalization~\cite{goodfellow2016deep}. Figure \ref{batch} (bottom) shows that even for a small block length ($L = 10$) when enumeration of all possible codes becomes possible, random mini-batching outperforms fixed full batch which enumerates all possible codes in each batch. Note that training with all possible codewords leads to a worse test performance, while training with large random batch (5000) outperforms the full batch settings. Thus we conclude that using a large random batch gives a better result. In this paper, due to GPU memory limitations, we use a batch size of 1000.

\subsection{Use Binary Crossentropy Loss}
\begin{figure}[!ht] 
\centering
\includegraphics[width=0.38\textwidth]{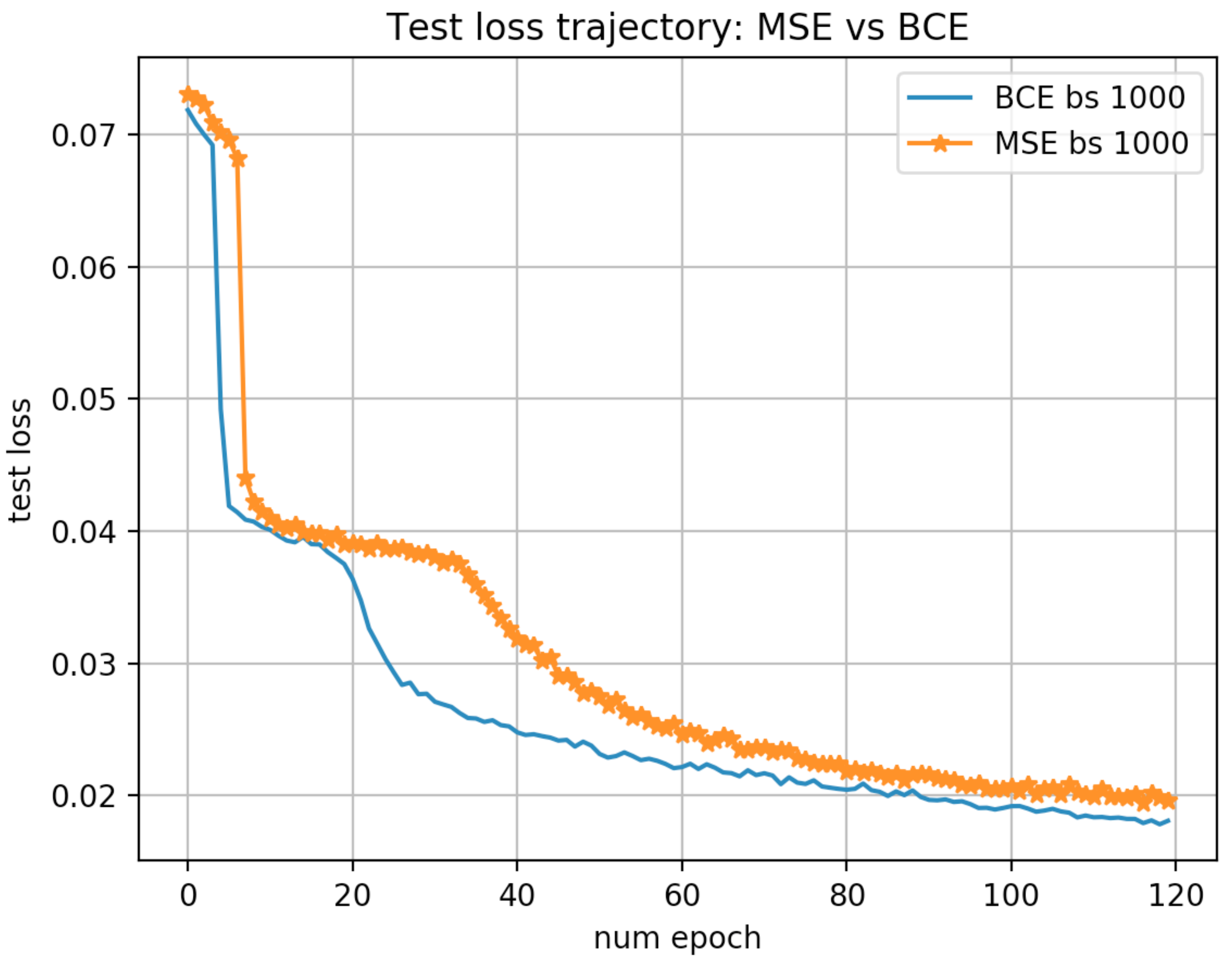}\ \ \ 
\includegraphics[width=0.40\textwidth]{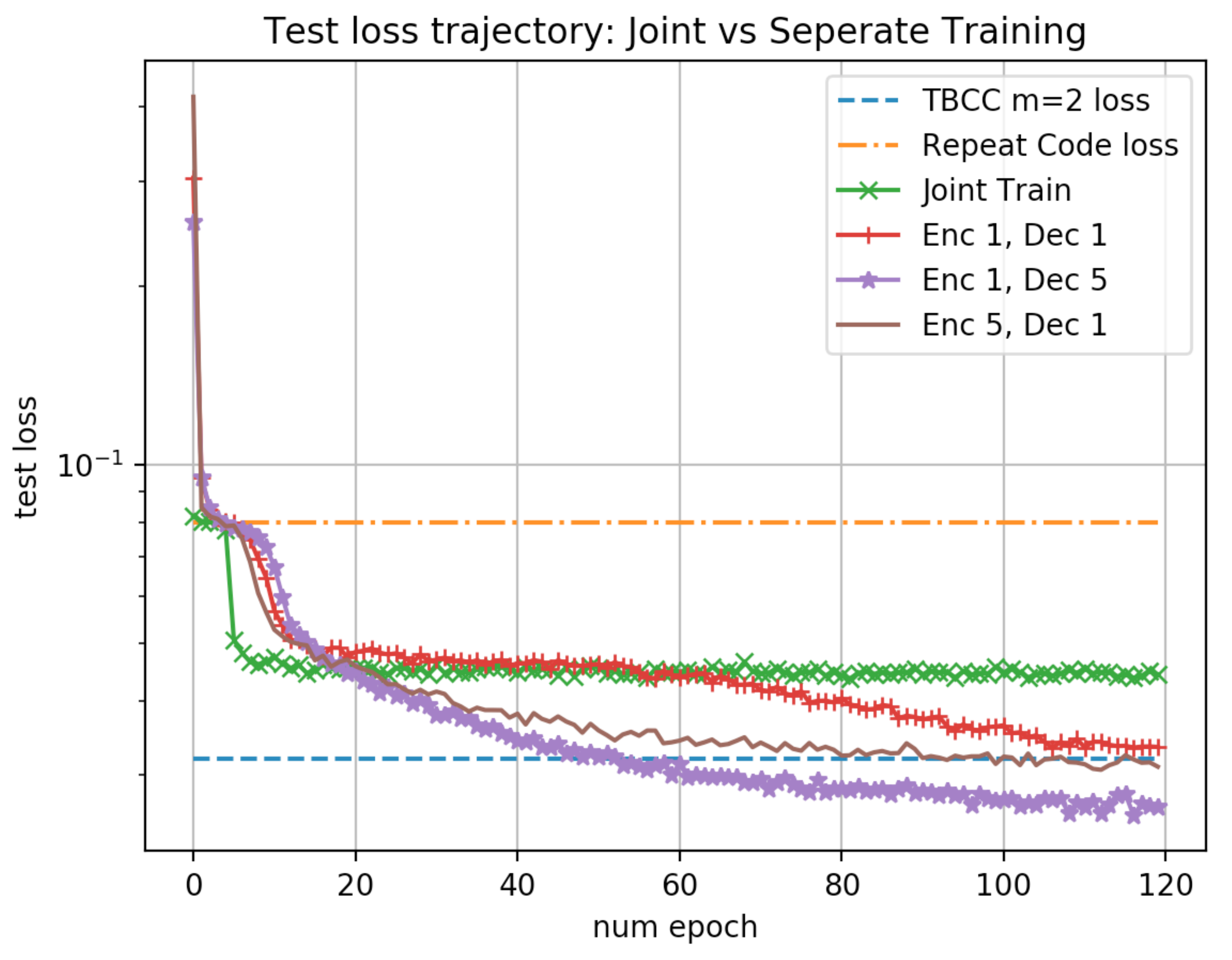}\ \ \ 
\caption{Test loss trajectory different loss functions (top), Training jointly vs separately (bottom)}\label{loss}
\end{figure}

The input $u \in \{0, 1\}^k$ and output $\hat u \in \{0, 1\}^k$ are binary digits, which makes training Channel AE a binary self-supervised classification problem \cite{o2016learning}. Binary Crossentropy (BCE) is better due to surrogate loss argument \cite{buja2005loss}. MSE and its variants can be used as the loss function for Channel AE \cite{o2017deep}\cite{gruber2017deep} as MSE offers implicit regularization on over-confidence of decoding. The comparison of MSE and BCE loss is shown in Figure \ref{loss} (top). 

Although both the final test loss for BCE and MSE  tend to converge, MSE leads to slower convergence. This is due to the fact that MSE actually punishes overconfident bits, which makes the learning gradient more sparse. As faster convergence and better generalization are desired, BCE is used as a primary loss function.

\subsection{Separately Training Encoder and Decoder}
Training encoder and decoder jointly with end-to-end back-propagation leads to saddle points. We argue that training Channel AE entails separately training encoder and decoder \cite{aoudia2018end}. The accurate gradient of the encoder can be computed when the decoder is optimal for a given encoder. Thus after training encoder,  training decoder until convergence will make the gradient of encoder more trustable. However, at every step training decoder till convergence is computationally expensive. Empirically we compare different training methods in Figure \ref{loss} (bottom). Training encoder and decoder jointly saturates easily. Training encoder once and training decoder 5 times shows the best performance and is used in this paper.

\subsection{Adding Minimum Distance Regularizer}
Naively optimizing Channel AE results in a paired local optimum: a pair of sub-optimal encoder and decoder can be locked in a saddle-point. Adding regularization to loss is a common method to escape local optima \cite{goodfellow2016deep}. Coding theory suggests that maximizing minimum distance between all possible input messages\cite{richardson2008modern} improves coding performance. However, since the number of all possible messages increases exponentially with respect to the code block length, computing loss for the largest minimum distance for long block code becomes prohibitive. 

\begin{figure}[!ht] 
\centering
\includegraphics[width=0.5\textwidth]{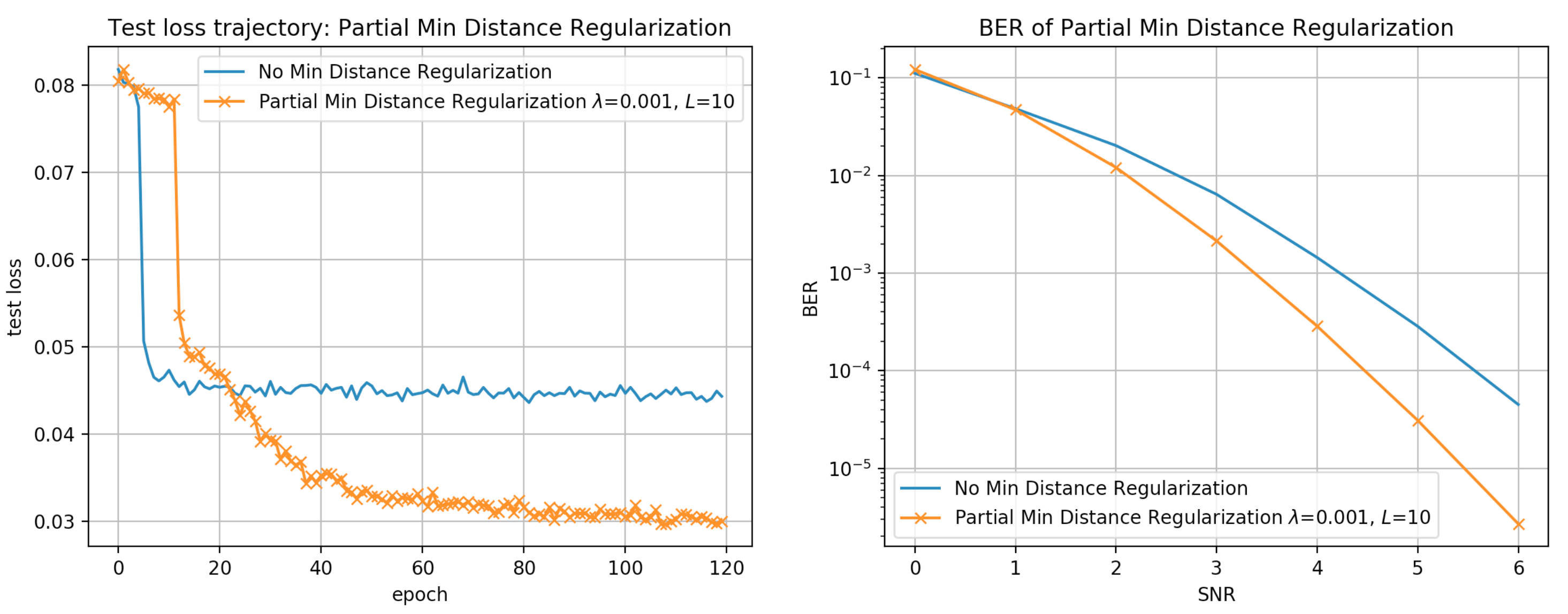}\ \ \ 
\includegraphics[width=0.25\textwidth]{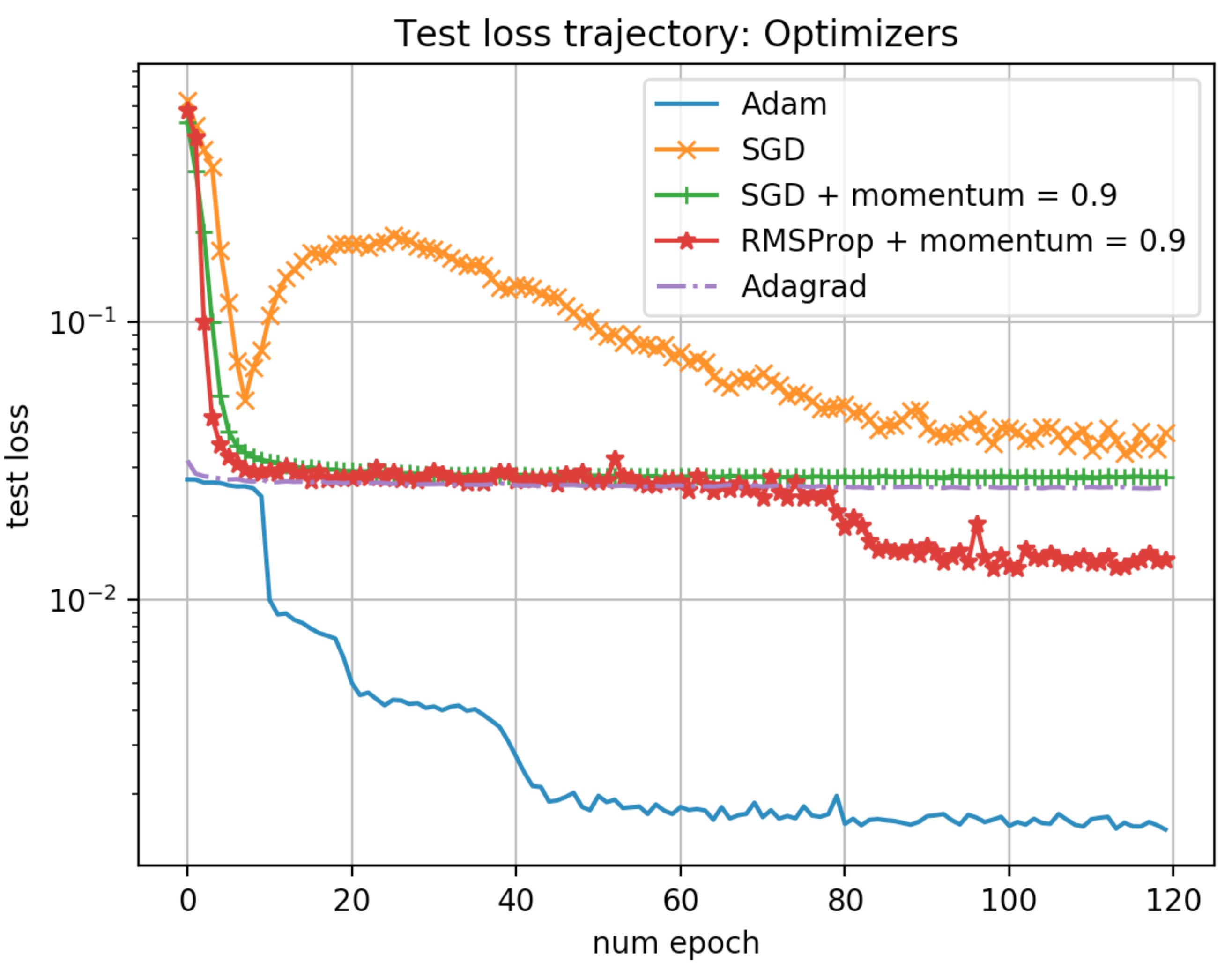}\ \ \ 
\caption{Effect of Partial Minimum Distance Regularization, Test loss trajectory (top left) and BER (top right). Test loss trajectory different optimizers with learning rate 0.001 (bottom)}\label{critic}
\end{figure}

Exploiting the locality inherent to RNN codes, we introduce a different loss term solely for the encoder which we refer to as the partial minimum code distance regularizer. Partial minimum code distance $d(u_L) = \min_{u_1, u_2 \in R^L, u_1 \neq u_2} \{||f_\theta(u_1) - f_\theta(u_2) ||_2^2 \}$ is the minimum distance among all message with length $L$. Computing pairwise distance requires $O({2^L \choose 2})$ computations. For a long block code with a block length $L_B>>L$, the RNN encoded portion $L$ has the minimum distance $M$, which guarantees that the block code with a block length $L_B$ has the minimum distance at least $M$, while the minimum distance can be as large as $\frac{L_B}{L}M$. Partial minimum code distance is a compromise over computation, which still guarantees large minimum distance under small block length $L$, while hoping the minimum distance on longer block length would still be large. The loss objective of Channel AE with partial maximized minimum code distance is $L(u) =E_{z}|| g_\phi(h(f_{\theta}(u)) + z) - u ||_2^2 + \lambda d(u_L)$.
To beat convolutional code via RNN autoencoder, we add minimum distance regularizer for block length 100, with $L=10$ and $\lambda = 0.001$. The performance is shown in Figure \ref{critic}, the top left panel shows the test loss trajectory, and the top right panel shows the BER. 

\subsection{Using Adam optimizer}
The learning rate for the encoder and the decoder has to be adaptive to compensate the realizations of noise with different magnitude. Also as training loss decreases, it is less likely to experience decoding error, which making the gradients for both encoder and decoder sparse. Adam\cite{adam2015kingma} has an adaptive learning rate and an exponential moving average for a sparse and non-stationary gradient, which in theory is a suitable optimizer. The comparison of different optimizers in Figure \ref{critic} (bottom) shows that Adam empirically outperforms all other optimizers, with faster convergence and better generalization. SGD fails to converge with learning rate 0.001 with high instability. Thus we use Adam for training Channel AE.

\subsection{Adding more capacity (parameters) to the decoder than the encoder}
Channel AE can be considered as an overcomplete autoencoder model with the noise injected in the middle layer. In what follows, we perform the analysis of the noise as done in [39] and for simplification consider only Minimum Square Error (MSE) (Binary Cross-entropy (BCE) loss would follow the same procedure). Assume $z\sim N(0, \sigma^2)$ is the added Gaussian noise, $u$ is the input binary bits, $f_\theta$ is the encoder, $g_\phi$ is the decoder, and $h(.)$ is the power constraint module. Applying Taylor expansion (see appendix for more details), the loss of Channel AE can be written as: $L = || g_\phi(h(f_{\theta}(u))) - u||_2^2 + \sigma^2||J_{g_\phi}(u)||_F^2$, where $||J_{g_\phi}(u)||_F^2 =\sum_{i}^k\sum_{j}^n (\frac{{\partial g_\phi(c)}_i}{\partial c_j})^2 $ is the Jacobian of function $g_\phi$. The reconstruction error $|| g_\phi(h(f_{\theta}(u))) - u||_2^2 $ can be interpreted as the coding error when no noise is added. With smaller $\frac{{\partial g_\phi(c)}_i}{\partial c_j}$, the decoder results in an invariance and robustness of the representation for small variations of the input~\cite{rifai2011contractive}, thus the Jacobian term $ \sigma^2 \sum_{i}^k\sum_{j}^n (\frac{{\partial g_\phi(c)}_i}{\partial c_j})^2 $ is the regularizer encouraging the decoder to be locally invariant to noise \cite{poole2014analyzing}. The Jacobian term reduces the sensitivity of the decoder, which improves the generalization of the decoders.

Empirically there exist an optimal training SNR for neural decoders \cite{kim2018communication} and Channel AE \cite{o2016learning}. When training with a large noise $\sigma^2$, the Jacobian term dominates, hence the reconstruction error becomes non-zero, which degrades the performance. When training with too small noise, the decoder is not local invariant, which reduces the generalization capability. 

Also as the Jacobian term only applies regularization to the decoder, the decoder needs more capacity. Empirically the neural decoder has to be more complicated than encoder. With training for 120 epochs, the encoder/decoder size 25/100 and 100/400 units shows an improved test loss, comparing to the cases where the encoder is less complicated than the decoder. As the encoder/decoder with 25/100 units works as well as encoder/decoder with size 100/400 units, we take 25 units encoder and 100 units decoder for most of our applications. Further hyperparameter optimization such as Bayesian Optimization\cite{snoek2012bo} could lead to an even better performance. 

\begin{figure}[!ht] 
\centering
{\tiny
\begin{tabular}{ |p{2cm}|p{2cm}|p{2cm}|  }
\hline
\multicolumn{3}{|c|}{Encoder and Decoder Hyperparameter Design after 120 epochs} \\
\hline
Enc Unit & Dec Unit &Test Loss\\
 \hline
 25   & 100    &0.180\\
 25   & 400    &0.640\\
 100   & 100    &0.690\\
 100   & 400    &0.181\\
 \hline
\end{tabular}\label{encdec}
}
\end{figure}


\section{Further Empirical Results}
\subsection{Alternative Minimum Distance Regularizer}
As RNN encoders have small dependency lengths (see section 4), small Hamming distance of messages may cause small code distance. Another method of regularizing is to directly regularize the minimum distance between the codewords for messages with small Hamming distances. The method starts by enumerating all messages within Hamming distance $s$ to a random message $u$ of length $L_B$, which contains $\sum_{i=1}^s {L_B \choose i} + 1$ messages, and then it computes the minimum distance among the enumerated messages as a regularization term. However, this method doesn't guarantee any minimum distance property even among short block length, and the computational complexity is high with even small $s$. Empirically this method doesn't work well.

\subsection{Loss Analysis for Encoder and Decoder Size}
This section shows the derivation of loss analysis used in the main text. Using Minimum Square Error (MSE), the loss of Channel AE is $L =E_{z}|| g_\phi(h(f_{\theta}(u)) + z) - u ||_2^2$. The output of the encoder is denoted as $c =h(f_{\theta}(u)) $. Using $1^{st}$ order Taylor expansion following~\cite{sabri2017effect}, with the assumption that noise $z$ is small and ignoring all higher order components, the decoder is approximated as: $ g_\phi(c + z) = g_\phi(c) + \frac{\partial g_\phi(c)}{\partial c}z + O(z)$

Note that the $1^{st}$ order Taylor expansion is a local approximation of a function. Hence, the assumption of ignoring higher order components is only valid with small $z$ locally. Then by expanding the MSE loss, we have:
$L = || g_\phi(h(f_{\theta}(u))) - u||_2^2 + E_{z} tr(z^T {\frac{\partial g_\phi(c)}{\partial c}}^T \frac{\partial g_\phi(c)}{\partial c} z )$, which yields: $L = || g_\phi(h(f_{\theta}(u))) - u||_2^2 + \sigma^2||J_{g_\phi}(u)||_F^2$.

\subsection{Low Latency Benchmark: Convolutional Code with Different Memory length}
\begin{figure*}[!ht] 
\centering
\includegraphics[width=0.9\textwidth]{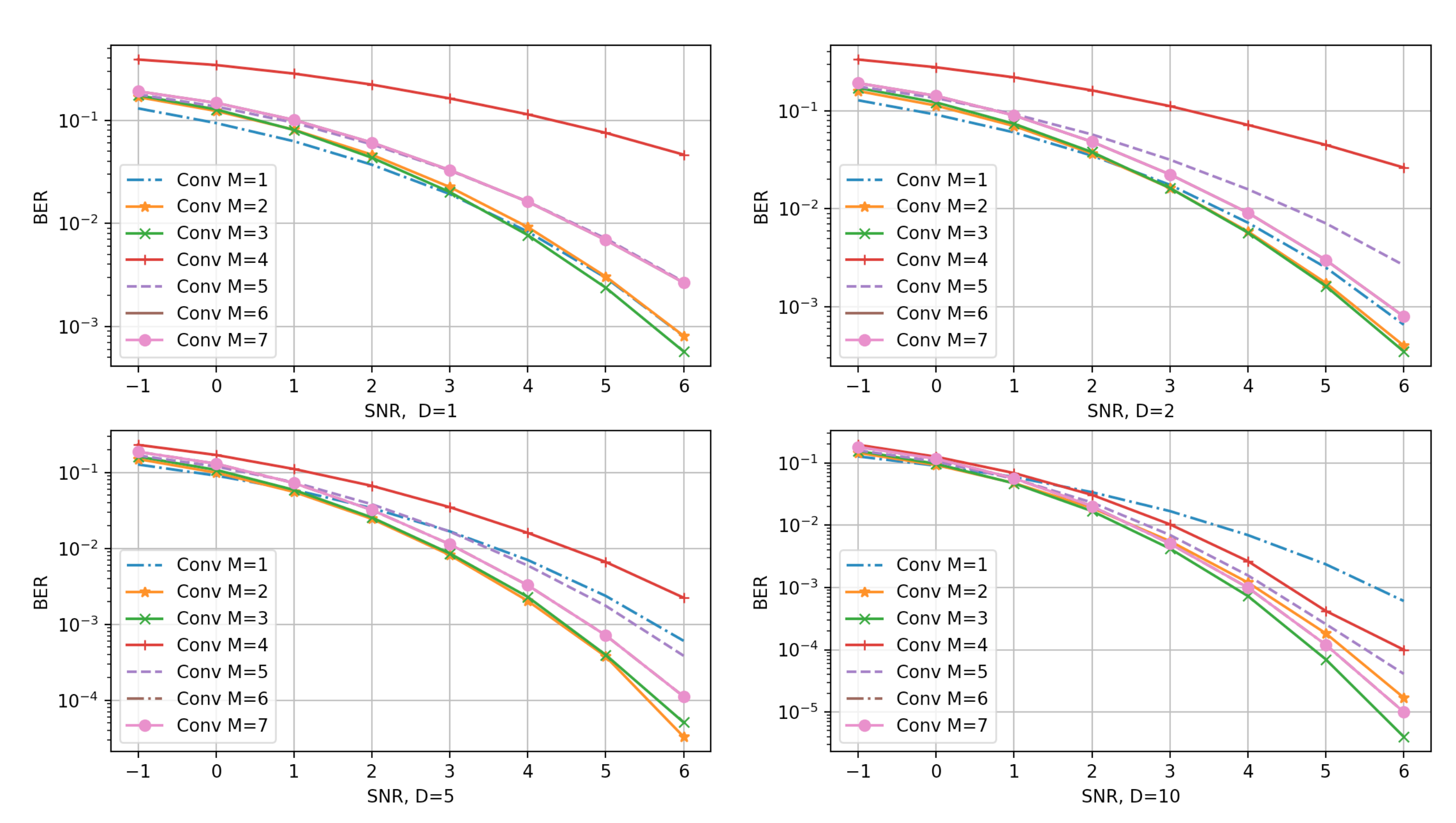}\ \ \ 
\vspace{2em}\\
\includegraphics[width=0.9\textwidth]{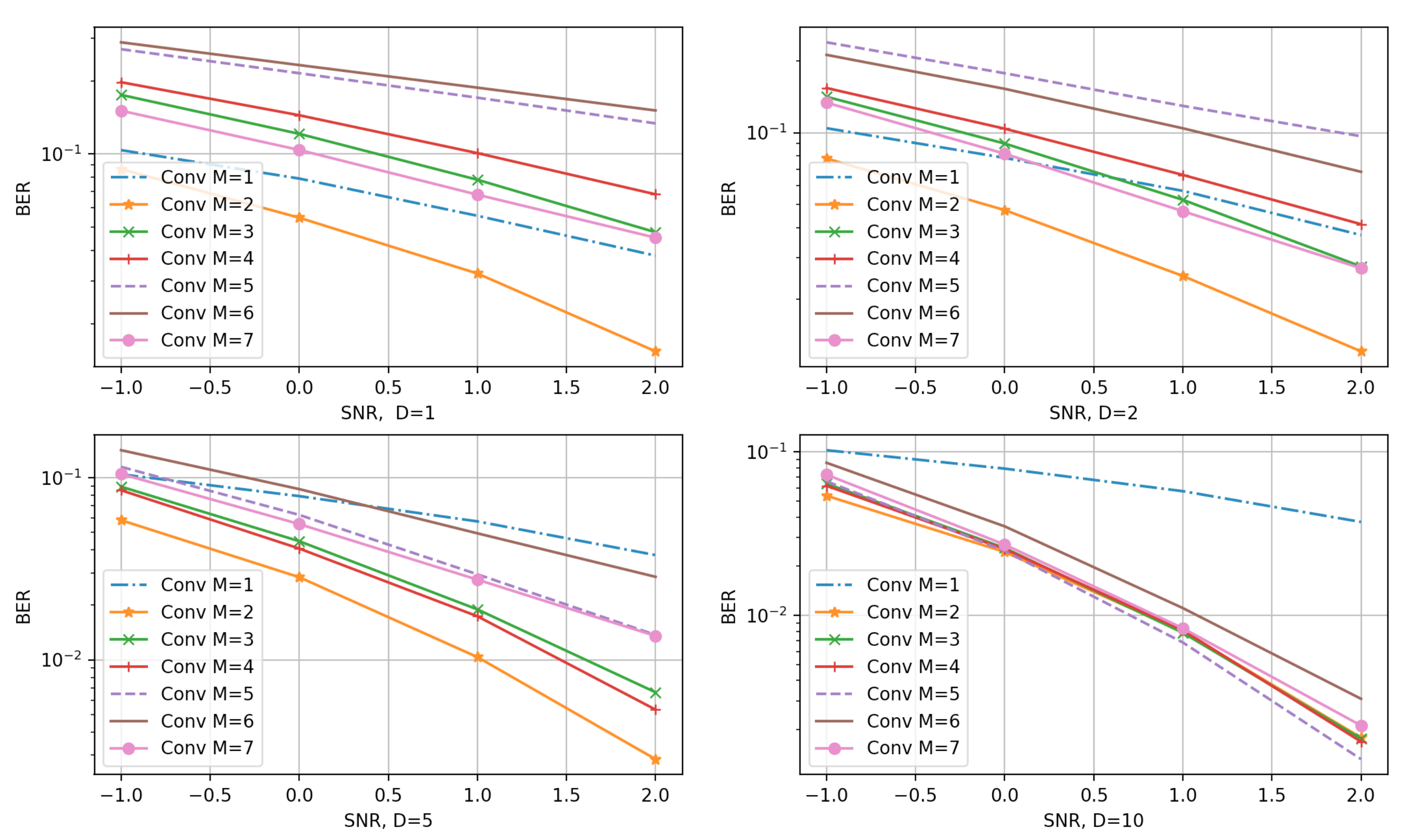}\ \ \
\caption{Convolutional Code with different delay constraint on rate 1/2 (top), rate 1/3 (bottom)}\label{ll_conv_bm}
\end{figure*}

The benchmarks of applying convolutional codes under low latency constraint are shown in Figure \ref{ll_conv_bm}. Note that there doesn't exist a single convolutional code that is universally best for various delay constraints and code rates. The convolutional codes reported in the main section are using the best convolutional codes shown in Figure \ref{ll_conv_bm}.

\end{document}